\newif\ifarxivtypeset
\arxivtypesettrue

\ifarxivtypeset
  \documentclass[utf8]{article}
  \usepackage{arxiv}
\else
  \documentclass[utf8]{frontiersSCNS} 
\fi

\usepackage[utf8]{inputenc} 
\usepackage[T1]{fontenc}    
\usepackage{url,hyperref,lineno,microtype,subcaption}
\usepackage{amsmath, amsthm, amssymb, amsfonts, mathtools, mathrsfs}
\usepackage[algo2e,ruled,algonl]{algorithm2e}
\usepackage[onehalfspacing]{setspace}
\usepackage{xcolor}

\ifarxivtypeset
\else
  \linenumbers
\fi

\def\keyFont{\fontsize{8}{11}\helveticabold }
\def\firstAuthorLast{Williams-Young {et~al.}} 
\def\Authors{David B. Williams-Young\,$^{1,*}$, Wibe A. de Jong$^1$, Hubertus J.J. van Dam$^2$ and Chao Yang$^1$}
\def\Address{$^1$Lawrence Berkeley National Laboratory, Computational Research Division,
Berkeley, CA, United States of America\\
$^2$Brookhaven National Laboratory, Computational Science Initiative, Upton, NY, United States of America}

\newif\ifexporttikz
\exporttikzfalse  
\ifexporttikz
    \usepackage{tikz}
    \usepackage{pgfplots}
    \pgfplotsset{compat=newest}
    \usepgfplotslibrary{external}
    \usetikzlibrary{patterns}
    \usetikzlibrary{intersections}
\else
    \usepackage{tikzexternal}
    \tikzsetexternalprefix{fig-export/}
\fi
\newcommand{\figname}[1]{\tikzsetnextfilename{#1}}

\definecolor{prota1}{HTML}{F5793A}
\definecolor{prota2}{HTML}{A95AA1}
\definecolor{prota3}{HTML}{85C0F9}
\definecolor{prota4}{HTML}{447490}

\usepackage[capitalize]{cleveref}
\crefname{figure}{Fig.}{Figs.}
\Crefname{figure}{Figure}{Figures}
\crefname{table}{Tab.}{Tabs.}
\Crefname{table}{Table}{Tables}
\crefname{equation}{Eq.}{Eqs.}
\Crefname{equation}{Equation}{Equations}
\crefname{section}{Sec.}{Secs.}
\Crefname{section}{Section}{Sections}
\crefname{paragraph}{Sec.}{Secs.}
\Crefname{paragraph}{Section}{Sections}
\crefname{algorithm}{Alg.}{Algs.}
\Crefname{algorithm}{Algorithm}{Algorithms}
\crefname{theorem}{Thm.}{Thms.}
\Crefname{theorem}{Theorem}{Theorems}
\crefname{corollary}{Cor.}{Cors.}
\Crefname{corollary}{Corollary}{Corollaries}

\newcommand{\densitySymb}[0]{\ensuremath{\rho}}
\newcommand{\density}[1]{\ensuremath{\densitySymb\left( #1 \right)}}
\newcommand{\densityR}[0]{ \density{\mathbf{r}} }

\newcommand{\bfn}[2]{\ensuremath{\phi_{#1}( #2 )}}
\newcommand{\bfnR}[1]{ \bfn{#1}{\mathbf{r}} }

\newcommand{\energy}[1]{\ensuremath{\mathcal{#1}}}

\begin{document}
\ifarxivtypeset
\else
  \onecolumn
  \firstpage{1}
\fi

\ifarxivtypeset
  \title{On the Efficient Evaluation of the Exchange Correlation Potential on Graphics Processing Unit Clusters} 
\else
  \title[Evaluation of the XC Potential on GPU Clusters]{On the Efficient Evaluation of the Exchange Correlation Potential on Graphics Processing Unit Clusters} 
\fi

\ifarxivtypeset
  \author{\Authors \\~\\~\\
  \Address\\~\\
  $^*$\texttt{dbwy@lbl.gov}} 
\else
  \author[\firstAuthorLast ]{\Authors} 
  \address{} 
  \correspondance{} 
  
  \extraAuth{}
\fi

\maketitle

\begin{abstract}

The predominance of Kohn-Sham density functional theory (KS-DFT) for the
theoretical treatment of large experimentally relevant systems in molecular
chemistry and materials science relies primarily on the existence of efficient
software implementations which are capable of leveraging the latest advances
in modern high performance computing (HPC). With recent trends in HPC leading
towards in increasing reliance on heterogeneous accelerator based
architectures such as graphics processing units (GPU), existing code bases
must embrace these architectural advances to maintain the high-levels of
performance which have come to be expected for these methods. In this work, we
purpose a three-level parallelism scheme for the distributed numerical
integration of the exchange-correlation (XC) potential in the Gaussian
basis set discretization of the Kohn-Sham equations on large computing
clusters consisting of multiple GPUs per compute node. In addition, we purpose
and demonstrate the efficacy of the use of batched kernels, including batched
level-3 BLAS operations, in achieving high-levels of performance on the GPU. We
demonstrate the performance and scalability of the implementation of the
purposed method in the NWChemEx software package by comparing to the existing 
scalable CPU XC integration in NWChem.

\ifarxivtypeset
\else
  \keyFont{\section{Keywords:} density functional theory, graphics processing unit, high-performance computing, parallel computing, quantum chemistry} 
\fi
\end{abstract}

\section{Introduction}

Kohn-Sham density functional theory (KS-DFT)
\cite{hohenberg64_inhomogeneous,kohn65_self} is unequivocally the computational
workhorse of theoretical chemistry and materials design. With the excellent
balance of its computational cost to its ability to accurately predict physical
phenomena, KS-DFT is nearly without equal in the routine theoretical treatment
of large, experimentally relevant systems
\cite{ratcliff17_challenges,wu19_density}.  A primary factor contributing to
the popularity of KS-DFT methods is the existence of highly optimized
and scalable software implementations capable of leveraging the latest advances
in modern high performance computing. The existence of such software enables
the treatment of increasingly larger and more complicated systems as computing
resources become large enough to accommodate them.  Historically, these
optimizations have amounted to considering the underlying details of
homogeneous computing platforms such as shared and distributed memory
multi-core central processing unit (CPU) architectures to exploit memory
hierarchies, distributed node topology and interconnection, and computing
features such as single-instruction multiple data (SIMD) instructions, fused
multiply-add (FMA), etc. 
\cite{belling99_quantum,lasinski08_optimization,deijong10_utilizing,nguyen17_automatic,blyaska17_performance,jacquelin17_towards,brown08_massively,petrone18_efficient} 
However, as we approach the exascale computing era, the
emergence of more heterogeneous computing architectures render non-trivial the
direct application of existing algorithms and code bases to target these
complex architectures. As such, for KS-DFT to remain relevant in
the age of exascale and post-exascale computing, methods developers must be
prepared to embrace these emerging architectures to maintain the high standard
of computational performance which has come to be expected.

In recent years, the trajectory of high-performance computing has lead to an
increasing reliance on the use accelerators, such as graphics processing units
(GPU), to perform the majority of the floating point operations (FLOPs) on new
and emerging computing resources \cite{kindratenko09_gpu,parnell19_trends}.
For a detailed treatise on the details and challenges presented by these and
other emerging architectures and their use in conjunction with electronic
structure calculations, we refer to the work of \cite{gordon20_novel}. In this
work, we limit our discussion to the optimization of KS-DFT methods on NVIDIA
GPUs (in particular the NVIDIA Tesla V100) using the Compute Unified Device
Architecture (CUDA) programming platform \cite{cook12_cuda}.

Recently, there has been significant research effort afforded to porting
electronic structure software to the GPU \cite{gordon20_novel}. In the case of
large scale calculations, much work has gone into the development of
massively parallel GPU implementations of methods based on plane wave
\cite{jia19_parallel,maintz11_speeding,wang11_large}, real-space
\cite{andrade13_real,hakala13_parallel}, finite-element \cite{das19_fast,motamarri20_dft}, and
various other discretizations
\cite{huhn20_gpu,genovese09_density,yoshikawa19_gpu,van16_gpu} of the Kohn-Sham
equations. In this work, we consider the Gaussian basis set discretization of
the Kohn-Sham equations \cite{pople92_kohn}, which poses a number of challenges
for GPU implementations. The majority of these challenges
revolve around the computation of molecular integrals over Gaussian basis
functions. Of the required integrals, the electron repulsion integrals (ERIs) and
the exchange-correlation (XC) potential are among the most costly and the most
challenging to port to GPU architectures. 
Over the years, there has been a considerable amount of research
devoted to porting implementations of Gaussian basis set KS-DFT to the GPU
\cite{kussmann17_employing,madushanka20_parallel,peters20_combining,luehr16_gaussian,yasuda08_accelerating,titov13_generating,brown10_a}; however, the vast majority
of this work has been centered around the evaluation and digestion of the ERIs
in the construction of the Fock matrix
\cite{kalinowski17_arbitrary,kussmann17_employing,ufimtsev08_quantum1,ufimtsev09_quantum2,ufimtsev09_quantum3,miao13_acceleration,asadchev10_696,laqua20_highly}. 
On the other hand, the XC potential has received much less treatment in the 
literature in this regard 
\cite{yasuda08_accelerating,madushanka20_parallel,luehr16_gaussian}. This disparity
is understandable due to the fact that for large systems, the
ERI related contributions to the Fock matrix are computationally dominant and
the most challenging to parallelize. However, with recent advances in
semi-numerical techniques for exact exchange which have shown great promise in
early GPU implementations \cite{laqua20_highly}, ERI dominated calculations are
quickly becoming computationally competitive with the evaluation of the XC
potential by current methods. Further, current accounts of GPU implementations
of the XC integration have been limited to the devices which are accessible
within a particular compute node. To the best of the authors' knowledge, there
does not exist a GPU accelerated distributed memory evaluation of the XC potential
using Gaussian basis sets as of this report. Thus, in this work, we propose a
three-level parallelism scheme for the scalable distributed evaluation of the
Gaussian basis XC potential on large clusters of GPUs. 

In general, there are a number of important features of GPU architectures 
one must consider in the development of high-performance software:
%
%
%
\begin{itemize}
  \item GPU architectures exhibit orders of magnitude more computational
  threads than CPU architectures, allowing for the expression of massive 
  concurrency within a single GPU device.

  \item The memory space which is directly accessible to GPU devices is much
  lower in capacity in comparison with their CPU counterparts (O(16GB-32GB)
  on the GPU in comparison to upwards of O(1TB) on the CPU).

  \item Memory access within device memory exhibits a much higher bandwidth
  than CPU memory (O(900 GB/s) on the GPU in comparison to O(20-50 GB/s)
  on the CPU). 

  \item Data transfers between host and device memory spaces are low bandwidth
  (O(80 GB/s) with advanced technologies such as NVLink, O(35 GB/s) over PCIe),
  thus data transfers often pose a non-trivial overhead in GPU applications which
  require movement of large volumes of data.
\end{itemize}
A consequence of these features is that, despite the large number of
threads which are available to the GPU to perform computation, data locality
must be carefully tuned to exploit the low capacity device memory as to allow
for the expression of concurrency but also to avoid high cost and inherently
serial data transfers between host and device.  As such, those algorithms which
are able to express massive concurrency on local data without being interrupted
by synchronization points such as data transfers and memory allocations are
typically the best suited for GPU application. A key aspect of the method
proposed in this report is the optimization of data movement within the XC
integration as to express massive concurrency using data which resides in device
memory without transfers between host and device. 

Scientific applications often rely on the
existence of highly tuned linear algebra libraries (such as vendor implementations of BLAS and LAPACK) to achieve high levels of performance on 
contemporary and emerging architectures \cite{dongerra98_numerical}.
Over the years, many areas of matrix computation have achieved significant
performance improvements through the use of GPU accelerators
\cite{lawn266,fatahalian2004understanding,herault2019generic}. However, unless
the matrix computations needed by a particular application are large enough
as to fully exploit the resources of the device, it is unlikely that single
matrix operation such as matrix-matrix multiplication will be able to achieve high 
computational occupancy on the device. An important achievement in high-performance numerical
linear algebra has been the advent of highly-tuned batched implementations of
commonly encountered matrix operations, such as matrix-matrix multiplication,
triangular factorization, etc
\cite{haidar15_batched,abdelfattah16_performance}. Such batched implementations are provided in both vendor tuned (such as cuBLAS and cuSOLVER provided by NVIDIA) and
open source (such as MAGMA \cite{tdb10,ntd10,tensors}) GPU accelerated linear algebra libraries. In these batched
implementations, efficiency is achieved by dramatically
increasing the throughput of the matrix operations via concurrent execution
within a single device. Thus, if an application requires the manipulation of
many small matrices in a manner which allows for concurrent execution (such as
KS-DFT), large performance improvements can be made by utilizing these batched
implementations (see e.g. \cite{motamarri20_dft}). GPU accelerated BLAS has 
previously been used in the context of XC computations \cite{yasuda08_accelerating}.
In this work, we examine the use of batched BLAS to further accelerate these
operations to improve overall time-to-solution. 

This work will be organized as follows. \Cref{sec:ksdft,sec:num_int} will
briefly review the pertinent theory and high-level algorithmic constructs
related to the XC integration. \Cref{sec:dist_xc_eval} will then describe the
proposed method for the scalable, three-level parallelism scheme for the
distributed XC integration on clusters of GPUs. \Cref{sec:results} will
demonstrate the performance and scalability of the purposed method in
comparison to an existing high-performance CPU implementation using a wide range
of molecules, basis sets and quadrature sizes. Finally, \cref{sec:conclusions}
will conclude this work and offer insight into the impact of the purposed
method and briefly discuss future research directions.

\section{Methods}

\subsection{Kohn-Sham Density Functional Theory}
\label{sec:ksdft}

In KS-DFT, the total electronic energy within a particular density functional
approximation (DFA) takes the form \cite{parr_dftbook}
\begin{equation}
\energy{E}^{tot} = \energy{T}_s + \energy{V}_{ne} + \energy{J} - c_x\energy{K} + 
                   \energy{E}^{xc}, \label{eq:etot}
\end{equation}
where $\energy{T}_s$ and $\energy{V}_{ne}$ are the (non-interacting) kinetic
and electron-nuclear attraction energies, and $\energy{J}$ and $\energy{K}$ are
the classical Coulomb and exact exchange energies, respectively.
$c_x\in\mathbb{R}$ is a parameter which scales the contribution of
exact-exchange to the electronic energy.
$c_x=0$ is used for ``pure" DFAs whereas DFAs which uses $c_x\neq 0$ are
referred to as ``hybrid" DFAs \cite{becke93_density}. Without loss of
generality in the following, we will take $c_x=0$, though we note that the
algorithms presented in the following sections may also be extended to hybrid
methods without modification.  $\energy{E}^{xc}$ is the exchange-correlation (XC)
energy which is taken to be a functional
of the electron density $\densitySymb : \mathbb{R}^3 \rightarrow \mathbb{R}$.
In this work, we restrict our discussion to spin-restricted DFAs within the
generalized gradient approximation (GGA) \cite{perdew86_accurate,perdew86_density}, 
i.e. $\energy{E}^{xc}$ is approximated to only depend on $\densitySymb$ and its
gradient $\nabla\densitySymb : \mathbb{R}^3 \rightarrow \mathbb{R}^3$.
We note for completeness that the information
presented in this and the following sections may be extended to both
spin-unrestricted and spin-generalized KS-DFT methods as well as more advanced
DFAs (such as the meta-GGA) 
with the addition
of only a few intermediates \cite{petrone18_efficient,egidi17_two}.
As $\nabla\densitySymb$ is a vector valued quantity, and thus dependent on
reference frame quantities such as molecular orientation, it is canonical to
express $\energy{E}^{xc}$ as
\begin{equation}
\energy{E}^{xc} = 
  \int_{\mathbb{R}^3} \varepsilon( \{U(\mathbf{r})\} )\densityR \,\mathrm{d}^3\mathbf{r},
  \label{eq:exc_vvar}
\end{equation}
where $\varepsilon$ is an energy-density which depends on a set of so-called ``U"-variables, $\{U(\mathbf{r})\}$,
which are independent of reference frame. Within the GGA, the canonical
choice for these variables are
$\{ U(\mathbf{r}) \} = \{ \densityR, \gamma(\mathbf{r}) \}$ with
$\gamma(\mathbf{r}) = \| \nabla\densityR \|$.

By expanding the density in a finite set of basis functions, 
$\mathcal{S} = \{\bfnR{\mu}\}_{\mu=1}^{N_b}$, 
\begin{equation}
\densityR = \sum_{\mu\nu} P_{\mu\nu} \bfnR{\mu} \bfnR{\nu},
\end{equation}
where $\mathbf{P}$ is the density matrix, the Kohn-Sham Fock matrix takes the form \cite{parr_dftbook}
\begin{equation}
  \mathbf{F} = \mathbf{h} + \mathbf{J} +
  \mathbf{V}^{xc}. \label{eq:ksfock}
\end{equation}
$\mathbf{h}$ is the basis representation of the density independent core
Hamiltonian (e.g. the sum of kinetic energy and external potential operators),
and $\mathbf{J}$ is the basis representations of the classical Coulomb operator. Note 
that we have dropped the exact exchange term in \cref{eq:etot} as we have taken
$c_x=0$.  $\mathbf{V}^{xc}$ is the XC potential which may be expressed as
\cite{petrone18_efficient,burow11_linear,yasuda08_accelerating}
\begin{equation}
V_{\mu\nu}^{xc} = 
  \int_{\mathbb{R}^3} 
    \bfnR{\mu} Z_{\nu}(\mathbf{r}) + Z_{\mu}(\mathbf{r}) \bfnR{\nu}
  \mathrm{d}^3\mathbf{r},
  \label{eq:vxc_exact}
\end{equation}
where
\begin{equation}
Z_\mu(\mathbf{r}) = 
  \frac{1}{2}\frac{\partial \varepsilon(\{U(\mathbf{r})\})}{\partial\densitySymb} \bfnR{\mu} + 
  2\frac{\partial \varepsilon(\{U(\mathbf{r})\})}{\partial\gamma}
    \nabla\densityR\cdot\nabla\bfnR{\mu}. \label{eq:zvec_r}
\end{equation}
Note that the partial derivatives of $\varepsilon$ are to be evaluated with
the U-variables calculated at argument of $Z_\mu$.

\Cref{eq:ksfock,eq:vxc_exact,eq:zvec_r} are general to any (real-valued) basis set expansion.
In this work, we consider atomically centered contracted Gaussian basis functions 
of the form 
\begin{equation}
\phi_\mu(\mathbf{r}) = (x-R_x)^l (y-R_y)^m (z-R_z)^n 
  \sum_{\xi=1}^{n^\mu_\xi} d^\mu_\xi 
    \exp\left( -\alpha^\mu_\xi (\mathbf{r} - \mathbf{R}_\mu)^2\right), 
  \label{eq:gau_cart}
\end{equation}
where $\mathbf{R}_\mu = \{R_x,R_y,R_z\}$, $n^\mu_\xi$ is the contraction depth, $d^\mu_\xi$ is a
contraction coefficient, and $L=l+m+n$ is the total angular momentum.  Each term in
the sum is referred to as a primitive Gaussian function. Contracted basis functions with the same $L$,
$\{d^\mu_\xi\}$, $\{\alpha^\mu_\xi\}$, and $\mathbf{R}_\mu$ will be referred to as a basis
shell. Functions of the form \cref{eq:gau_cart} are referred to as Cartesian
Gaussian functions, and each Cartesian shell with angular momentum $L$ consists
of $L(L+1)$ functions. For $L>1$, there is often a linear dependency among the
functions within each Cartesian shell which may be addressed by transforming these shells
to a set of spherical Gaussian functions \cite{schlegel95_transformation}. 
Each spherical Gaussian shell consists of $2L+1$ linearly independent functions.
Not all Gaussian basis sets which consist of functions with $L>1$ require
this transformation to be linearly independent, and we will note when such
a transformation has taken place.

\ifarxivtypeset
\else
  ~\\  
\fi
\subsection{Numerical Integration of Molecular Integrands}
\label{sec:num_int}

Even for the simplest forms of $\varepsilon$, neither \cref{eq:exc_vvar} nor
\cref{eq:vxc_exact} admit analytic expressions, thus these integrations must be
performed numerically. For molecular integrands, i.e. integrands with
non-trivial behavior in the vicinity of atomic nuclei in polyatomic systems, a
particularly attractive approach is to perform the numerical integration as a
sum over weighted atomic integrands \cite{becke88_a}. For a molecular integrand
$f:\mathbb{R}^3 \rightarrow \mathbb{R}$, we may decompose its integral over
$\mathbb{R}^3$ as
\begin{equation}
\int_{\mathbb{R}^3} f(\mathbf{r})\,\mathrm{d}^3\mathbf{r} = 
\sum_{A=1}^{N_A} I_A[f], \qquad
I_A[f] = \int_{\mathbb{R}^3} p_A(\mathbf{r}) f(\mathbf{r})\,\mathrm{d}^3\mathbf{r},
\label{eq:atomic_partition}
\end{equation}
where $N_A$ is the number of atoms, and $p_A:\mathbb{R}^3\rightarrow
\mathbb{R}$ is an atomic partition function which obeys $\sum_A p_A(\mathbf{r})
= 1$, $\forall \mathbf{r}\in\mathbb{R}^3$. Each atomic integrand $I_A[f]$ may
then be approximated by a quadrature rule
\begin{equation}
I_A[f] \approx \sum_{i\in\mathcal{Q}_A} w^A_i f(\mathbf{r}^A_i), 
\quad w^A_i = p^A(\mathbf{r}^A_i)w^q_i \label{eq:num_int}
\end{equation}
where $\mathcal{Q}_A=\{(w^A_i, \mathbf{r}^A_i)\}_{i=1}^{N^A_g}$ is a set of
quadrature points indexed by $i$ centered around the $A$-th nucleus with
atomically-scaled quadrature weights $w^A_i$. $\{w_i^q\}$ is the set of
unmodified weights associated with the base quadrature around a particular
nucleus. For convenience in the following, we define the total quadrature 
\begin{equation*}
\mathcal{Q} = \bigcup_{A} \mathcal{Q}_A = \{(w_i, \mathbf{r}_i)\}_{i=1}^{N_g},
\end{equation*}
where $N_g = \sum_A N_g^A$ is the total number of grid points needed to perform the 
numerical integration over the molecular integrand. Note that $w_i$ is assumed
to have the proper atomic scaling per \cref{eq:num_int}.

There are many possible choices for both the atomic partitioning
scheme
\cite{becke88_a,stratmann96_achieving,apra20_nwchem,laqua18_an} and base
quadratures around each atomic center
\cite{becke88_a,murray93_quadrature,mura96_improved,apra20_nwchem,treutler95_efficient,gill03_radial}.
In this work, we will use the following:
\begin{itemize}
  \item For the atomic partition function, we will use the scheme proposed by
  of Stratmann, Scuseria and Frisch (SSF) \cite{stratmann96_achieving}.
  \item For the base atomic quadrature, we will a spherical product grid consisting
  of the Mura-Knowles (MK) quadrature \cite{mura96_improved} for the radial 
  integration and the Lebedev-Laikov quadrature \cite{lebedev76_quadratures} 
  for the angular integration.
\end{itemize}
These schemes are chosen in part for the simplicity and robustness, as well as
their standard use in industry KS-DFT software.  Further, while it is standard
practice to perform angular grid pruning to reduce the number of grid points in
these product quadratures \cite{gill93_a,chien06_sg0,laqua18_an}, we perform
no such procedure here. We note that the methodological details presented
in this work are largely independent of such choices.

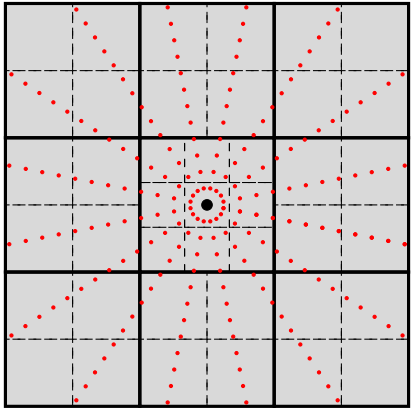
\begin{figure}[t]
  \figname{grid-partition}
  \centering
  \begin{minipage}{0.5\textwidth}
    \centering
    \begin{tikzpicture}

\begin{axis} [
  width=\textwidth,
  axis x line=center, 
  axis line style = {draw = none},
  axis y line=none, 
  ymin=-3.1, ymax=3.1,
  xmin=-3.1, xmax=3.1,
  axis equal image,
  ticks= none,
  xticklabels={}
]

\draw[fill=black!15] (-3,-3) rectangle (3,3);

\draw[fill=none,very thick] (-3,-3) rectangle (-1,-1);
\draw[fill=none,very thick] (-3,-1) rectangle (-1, 1);
\draw[fill=none,very thick] (-3, 1) rectangle (-1, 3);

\draw[fill=none,very thick] (-1,-3) rectangle ( 1,-1);
\draw[fill=none,very thick] (-1,-1) rectangle ( 1, 1);
\draw[fill=none,very thick] (-1, 1) rectangle ( 1, 3);

\draw[fill=none,very thick] ( 1,-3) rectangle ( 3,-1);
\draw[fill=none,very thick] ( 1,-1) rectangle ( 3, 1);
\draw[fill=none,very thick] ( 1, 1) rectangle ( 3, 3);

\draw[fill=none,dashed] (-1,     -1) rectangle (-1/3, -1/3);
\draw[fill=none,dashed] (-1,   -1/3) rectangle (-1/3,  1/3);
\draw[fill=none,dashed] (-1,    1/3) rectangle (-1/3,    1);
\draw[fill=none,dashed] (-1/3,   -1) rectangle ( 1/3, -1/3);
\draw[fill=none,dashed] (-1/3, -1/3) rectangle ( 1/3,  1/3);
\draw[fill=none,dashed] (-1/3,  1/3) rectangle ( 1/3,    1);
\draw[fill=none,dashed] ( 1/3,   -1) rectangle (   1, -1/3);
\draw[fill=none,dashed] ( 1/3, -1/3) rectangle (   1,  1/3);
\draw[fill=none,dashed] ( 1/3,  1/3) rectangle (   1,    1);

\draw[fill=none,dashed] (-3,-3) rectangle (-2,-2);
\draw[fill=none,dashed] (-3,-2) rectangle (-2,-1);
\draw[fill=none,dashed] (-2,-3) rectangle (-1,-2);
\draw[fill=none,dashed] (-2,-2) rectangle (-1,-1);

\draw[fill=none,dashed] (-3,-1) rectangle (-2, 0);
\draw[fill=none,dashed] (-3, 0) rectangle (-2, 1);
\draw[fill=none,dashed] (-2,-1) rectangle (-1, 0);
\draw[fill=none,dashed] (-2, 0) rectangle (-1, 1);

\draw[fill=none,dashed] (-3, 1) rectangle (-2, 2);
\draw[fill=none,dashed] (-3, 2) rectangle (-2, 3);
\draw[fill=none,dashed] (-2, 1) rectangle (-1, 2);
\draw[fill=none,dashed] (-2, 2) rectangle (-1, 3);

\draw[fill=none,dashed] (-1,-3) rectangle ( 0,-2);
\draw[fill=none,dashed] (-1,-2) rectangle ( 0,-1);
\draw[fill=none,dashed] ( 0,-3) rectangle ( 1,-2);
\draw[fill=none,dashed] ( 0,-2) rectangle ( 1,-1);

\draw[fill=none,dashed] (-1, 1) rectangle ( 0, 2);
\draw[fill=none,dashed] (-1, 2) rectangle ( 0, 3);
\draw[fill=none,dashed] ( 0, 1) rectangle ( 1, 2);
\draw[fill=none,dashed] ( 0, 2) rectangle ( 1, 3);

\draw[fill=none,dashed] (1,-3) rectangle (2,-2);
\draw[fill=none,dashed] (1,-2) rectangle (2,-1);
\draw[fill=none,dashed] (2,-3) rectangle (3,-2);
\draw[fill=none,dashed] (2,-2) rectangle (3,-1);

\draw[fill=none,dashed] (1,-1) rectangle (2, 0);
\draw[fill=none,dashed] (1, 0) rectangle (2, 1);
\draw[fill=none,dashed] (2,-1) rectangle (3, 0);
\draw[fill=none,dashed] (2, 0) rectangle (3, 1);

\draw[fill=none,dashed] (1, 1) rectangle (2, 2);
\draw[fill=none,dashed] (1, 2) rectangle (2, 3);
\draw[fill=none,dashed] (2, 1) rectangle (3, 2);
\draw[fill=none,dashed] (2, 2) rectangle (3, 3);

\addplot[only marks, mark=*] table [ ] {
  0 0
};

\addplot[red, only marks, mark=*, mark options={scale=0.3}] table [ 
  x expr = \thisrowno{0}*0.25,
  y expr = \thisrowno{1}*0.25,
] {fig-src/data/unit-circle.dat};

\addplot[red, only marks, mark=*, mark options={scale=0.3}] table [ 
  x expr = \thisrowno{0}*0.5,
  y expr = \thisrowno{1}*0.5,
] {fig-src/data/unit-circle.dat};

\addplot[red, only marks, mark=*, mark options={scale=0.3}] table [ 
  x expr = \thisrowno{0}*0.75,
  y expr = \thisrowno{1}*0.75,
] {fig-src/data/unit-circle.dat};

\addplot[red, only marks, mark=*, mark options={scale=0.3}] table [ 
  x expr = \thisrowno{0},
  y expr = \thisrowno{1},
] {fig-src/data/unit-circle.dat};

\addplot[red, only marks, mark=*, mark options={scale=0.3},
         restrict x to domain=-3:3,
         restrict x to domain=-3:3
] table [ 
  x expr = \thisrowno{0}*1.25,
  y expr = \thisrowno{1}*1.25,
] {fig-src/data/unit-circle.dat};
\addplot[red, only marks, mark=*, mark options={scale=0.3},
         restrict x to domain=-3:3,
         restrict x to domain=-3:3
] table [ 
  x expr = \thisrowno{0}*1.5,
  y expr = \thisrowno{1}*1.5,
] {fig-src/data/unit-circle.dat};
\addplot[red, only marks, mark=*, mark options={scale=0.3},
         restrict x to domain=-3:3,
         restrict x to domain=-3:3
] table [ 
  x expr = \thisrowno{0}*1.75,
  y expr = \thisrowno{1}*1.75,
] {fig-src/data/unit-circle.dat};
\addplot[red, only marks, mark=*, mark options={scale=0.3},
         restrict x to domain=-3:3,
         restrict x to domain=-3:3
] table [ 
  x expr = \thisrowno{0}*2,
  y expr = \thisrowno{1}*2,
] {fig-src/data/unit-circle.dat};

\addplot[red, only marks, mark=*, mark options={scale=0.3},
         restrict x to domain=-3:3,
         restrict x to domain=-3:3
] table [ 
  x expr = \thisrowno{0}*2.25,
  y expr = \thisrowno{1}*2.25,
] {fig-src/data/unit-circle.dat};
\addplot[red, only marks, mark=*, mark options={scale=0.3},
         restrict x to domain=-3:3,
         restrict x to domain=-3:3
] table [ 
  x expr = \thisrowno{0}*2.5,
  y expr = \thisrowno{1}*2.5,
] {fig-src/data/unit-circle.dat};
\addplot[red, only marks, mark=*, mark options={scale=0.3},
         restrict x to domain=-3:3,
         restrict x to domain=-3:3
] table [ 
  x expr = \thisrowno{0}*2.75,
  y expr = \thisrowno{1}*2.75,
] {fig-src/data/unit-circle.dat};
\addplot[red, only marks, mark=*, mark options={scale=0.3},
         restrict x to domain=-3:3,
         restrict x to domain=-3:3
] table [ 
  x expr = \thisrowno{0}*3,
  y expr = \thisrowno{1}*3,
] {fig-src/data/unit-circle.dat};

\addplot[red, only marks, mark=*, mark options={scale=0.3},
         restrict x to domain=-3:3,
         restrict x to domain=-3:3
] table [ 
  x expr = \thisrowno{0}*3.25,
  y expr = \thisrowno{1}*3.25,
] {fig-src/data/unit-circle.dat};
\addplot[red, only marks, mark=*, mark options={scale=0.3},
         restrict x to domain=-3:3,
         restrict x to domain=-3:3
] table [ 
  x expr = \thisrowno{0}*3.5,
  y expr = \thisrowno{1}*3.5,
] {fig-src/data/unit-circle.dat};
\addplot[red, only marks, mark=*, mark options={scale=0.3},
         restrict x to domain=-3:3,
         restrict x to domain=-3:3
] table [ 
  x expr = \thisrowno{0}*3.75,
  y expr = \thisrowno{1}*3.75,
] {fig-src/data/unit-circle.dat};
\addplot[red, only marks, mark=*, mark options={scale=0.3},
         restrict x to domain=-3:3,
         restrict x to domain=-3:3
] table [ 
  x expr = \thisrowno{0}*4,
  y expr = \thisrowno{1}*4,
] {fig-src/data/unit-circle.dat};

\end{axis}

\end{tikzpicture}
  \end{minipage}
  \caption{2-D cross-section of the grid batching scheme used in this work.
  The large black dot represents an atomic center and the small red dots
  represent quadrature points for spherical integration.  Thick solid lines
  represent the initial cuboid partition, and dashed lines represent the next
  partition level. Atomic centered cuboids are partitioned into 27 cubical
  domains while off-center cuboids are partitioned into octants.}
  \label{fig:grid-partition}
\end{figure}

It is well known that a naive application of
\cref{eq:atomic_partition,eq:num_int} to evaluate $\mathbf{V}^{xc}$ and
$\mathcal{E}^{xc}$ is very inefficient \cite{stratmann96_achieving}. This is due
to the fact that while Gaussian functions of the form \cref{eq:gau_cart} do not
admit compact support, their exponential character yields numerically
negligible contributions when evaluated far from their center. As such,
Gaussians of this form may be approximated to have compact support on a sphere
centered at their $\mathbf{R}_\mu$ with cutoff radius \cite{burow11_linear}
\begin{equation}
r^{cut}_\mu = \max_\xi \sqrt{\frac{1}{\alpha^\mu_\xi}\left(\frac{\ln\alpha^\mu_\xi}{2} - \ln \eta\right)},
  \label{eq:rcut}
\end{equation}
where $\eta$ is a tolerance for which $|\phi_\mu|<\eta$ for all points outside of
the sphere. In this work, we have chosen $\eta=10^{-10}$. Remark that the
cutoff radius only depends on the exponential coefficients, and thus may be
calculated at the level of basis shell rather than individual functions for
$L>0$. 
Given this cutoff criteria, one may form a list of basis shells which are
non-negligible for each quadrature point.  Rather than check each individual
quadrature points against $r^{cut}$ for each basis shell's cutoff radius, it is
canonical to group quadrature points which are spatially close into batches and
perform the coarse-grained screening for non-negligible basis shells at the
batch level rather than the quadrature points themselves.  This procedure is
known as micro-batching \cite{stratmann96_achieving} and is one of the primary
mechanisms by which linear scaling (with respect to system size) is achieved in
the evaluation of the XC potential. There are several ways to obtain the
quadrature batches
\cite{stratmann96_achieving,burow11_linear,madushanka20_parallel}. 
In this work, we recursively subdivide the domain spanned by the quadrature
points into cuboids until the number of quadrature points within each cuboid is
below a certain threshold. In this work, we have chosen this threshold to be
512 quadrature points.  In practice, this partitioning scheme produces batches
similar to the octree method of \cite{madushanka20_parallel}. However, rather
than bisecting every domain into octants, cuboids which contain an atomic
center are partitioned into 27 cuboids as shown in \cref{fig:grid-partition}.
Our experiments show that this procedure produces fewer batches with the same
non-negligible shell list which in turn improves the performance of the load
balancing scheme discussed later in this section.  However, much like the
choice of atomic quadrature and partition functions, the choice of batching
scheme does not effect the methodological details presented in this work just
as long as the batches produced are able to produce sufficiently short lists of
non-negligible basis shells.  For a total quadrature $\mathcal{Q}$, we denote
the set of quadrature batches produced by this procedure as 
$\mathcal{B} = \{\mathcal{B}_j\}$ such that
\begin{equation}
\mathcal{Q} = \bigcup_{\mathcal{B}_j \in \mathcal{B}} \mathcal{B}_j, \quad
\text{s.t.} \quad \mathcal{B}_j \cap \mathcal{B}_k = \emptyset \text{, for } j \neq k.
\end{equation}
In the case where the batches are defined by non-overlapping
cuboids surrounding an atomic center, basis shell screening may be accomplished by 
calculating the point of closest approach between the cuboid defining the batch
and the spheres defined by center $\mathbf{R}_\mu$ and radius $r^{cut}_\mu$ \cite{arvo13_graphics}. A description
of this procedure is given in \cref{alg:cube_sphere_intersection}.
For $\mathcal{B}_j \in \mathcal{B}$, we define the list of non-negligible basis
functions for $\mathcal{B}_j$ as $\mathcal{S}_j$, the number of non-negligible
basis functions as $N_b^j = |\mathcal{S}_j|$, and the number of quadrature points
in the batch as $N_g^j = |\mathcal{B}_j| $.

\begin{algorithm2e}[t]
  \caption{Basis Shell Screening via Cuboid-Sphere Intersection}
  \label{alg:cube_sphere_intersection}
  \SetKwInOut{kinput}{Input}
  \SetKwInOut{koutput}{Output}

  \BlankLine
  \kinput{ Sphere center $\mathbf{R}_\mu = \{R_x,R_y,R_z\}$, sphere radius $r^{cut}_\mu$, minimum (maximum)
  vertex defining the cuboid $V=\{V_x,V_y,V_z\}$ ($W=\{W_x,W_y,W_z\}$).}
  \BlankLine
  \koutput{ \textbf{True} if the cuboid and sphere spatially intersect, \textbf{False} otherwise.}

  \DontPrintSemicolon
  \BlankLine
  $d \leftarrow \left(r_\mu^{cut}\right)^2$ \\
  \BlankLine
  \For{ $p \in \{x,y,z\}$ } {
    \lIf{ $R_p < V_p$ } {
      $d \leftarrow d - (R_p - V_p)^2$
    } 
    \lElseIf{ $R_p > W_p$ } {
      $d \leftarrow d - (R_p - W_o)^2$
    }
  }
  \BlankLine
  \Return $(d < 0)$
\end{algorithm2e}

Another advantage of quadrature batching is the ability to cast the
evaluation of $\mathbf{V}^{xc}$ and $\mathcal{E}^{xc}$ in terms of efficient level-1 BLAS
operations such as dot products (DOT) and level-3 BLAS operations such as
matrix-matrix multiplication (GEMM) and symmetric rank-2K updates (SYR2K).  For
a particular batch $\mathcal{B}_j$, we may define a 
batch collocation matrix ($\boldsymbol{\Phi}^j$) and a local density matrix
($\mathbf{P}^j$)  as 
\begin{subequations}
\begin{align}
&\Phi_{\mu i}^j = \begin{cases}
\bfn{\mu}{\mathbf{r}_i}, & \text{for } i \in \mathcal{B}_j\text{ and }\mu \in \mathcal{S}_j \\
0, & \text{otherwise.}
\end{cases} \label{eq:local_basis_def}\\
&P_{\mu\nu}^j = \begin{cases}
P_{\mu\nu}, & \text{for } \mu,\nu \in \mathcal{S}_j \\
0, & \text{otherwise.}
\end{cases} \label{eq:local_density_def}
\end{align}
\label{eq:batch_local_mats}
\end{subequations}
In the following, we will refer to the extent to which $\boldsymbol{\Phi}^j$
and $\mathbf{P}^j$ are numerically zero due to basis function screening as
their local sparsity. This yields the following expressions for the density and
its gradient evaluated on the quadrature points within $\mathcal{B}_j$,
\begin{alignat}{3}
&\densitySymb_i^j       &&= \sum_{\mu \in \mathcal{S}_j} \Phi^j_{\mu i} X^j_{\mu i},        \quad &&\text{(DOT)} \label{eq:eval_local_den}\\
&\nabla\densitySymb_i^j &&= 2\sum_{\mu \in \mathcal{S}_j} \nabla\Phi^j_{\mu i} X^j_{\mu i}, \quad &&\text{(DOT)} \label{eq:eval_local_den_grad}\\
&\mathbf{X}^j           &&= \mathbf{P}^j \boldsymbol{\Phi}^j.                               \quad &&\text{(GEMM)} \label{eq:x_mat}
\end{alignat}
It should be understood from the context that the free index $i$ is restricted to quadrature points in
$\mathcal{B}_j$. Given these expressions, we may now express the XC related quantities as \cite{petrone18_efficient}
\begin{alignat}{3}
&\mathcal{E}^{xc} &&= \sum_{\mathcal{B}_j\in\mathcal{B}} \sum_{i \in \mathcal{B}_j} \varepsilon^j_i \densitySymb_i^j, \quad && \text{(DOT)}  \label{eq:exc_batch}\\
&V^{xc}_{\mu\nu}  &&= \sum_{\mathcal{B}_j\in\mathcal{B}} V_{\mu\nu}^j, && \label{eq:vxc_inc}\\ 
&\mathbf{V}^{j}   &&=  \mathbf{Z}^j \boldsymbol{\Phi}^{j,T} + \boldsymbol{\Phi}^j \mathbf{Z}^{j,T} , \quad && \text{(SYR2K)}
  \label{eq:vxc_batch}
\end{alignat}
with
\begin{align}
&\varepsilon^j_i = w_i\varepsilon(\{U(\mathbf{r}_i)\}), 
  \quad \frac{\partial\varepsilon^j_i}{\partial \rho} = w_i\frac{\partial \varepsilon(\{U(\mathbf{r}_i)\})}{\partial \rho} 
  \quad \frac{\partial\varepsilon^j_i}{\partial \gamma} = w_i\frac{\partial \varepsilon(\{U(\mathbf{r}_i)\})}{\partial \gamma},  \label{eq:eval_xc_batch}\\
&Z_{\mu i}^j   = \frac{1}{2} \frac{\partial\varepsilon^j_i}{\partial \rho} \Phi^j_{\mu i} + 
  2\frac{\partial\varepsilon^j_i}{\partial \gamma} \left(\nabla\densitySymb_i^j \cdot \nabla \Phi^j_{\mu i}\right). \label{eq:z_mat}
\end{align}
For brevity in the following, we define for $i \in \mathcal{B}_j$
\begin{equation}
  \boldsymbol{\densitySymb}^j = \left\lbrace \densitySymb_i^j \right\rbrace, \quad
  \nabla\boldsymbol{\densitySymb}^j = \left\lbrace \nabla\densitySymb_i^j \right\rbrace, \quad
  \boldsymbol{\varepsilon}^j = \left\lbrace \varepsilon_i^j \right\rbrace, \quad
  \boldsymbol{\varepsilon}_\rho^j   = \left\lbrace \frac{\partial \varepsilon_i^j}{\partial \rho} \right\rbrace, \quad
  \boldsymbol{\varepsilon}_\gamma^j = \left\lbrace \frac{\partial \varepsilon_i^j}{\partial \gamma} \right\rbrace.
\end{equation}

\begin{figure}[t]
  \figname{mat_compress}
  \centering
  \begin{minipage}{0.6\textwidth}
    \centering
    \begin{tikzpicture}

\begin{axis} [
  width=\textwidth,
  axis x line=center, 
  axis line style = {draw = none},
  axis y line=none, 
  ymin=-5.1, ymax=5.1,
  xmin=-10.1, xmax=11.1,
  axis equal image,
  ticks= none,
  xticklabels={}
]

\draw[] (-10,-5) rectangle (0,5);

\draw[line width=2pt,fill=prota1]   (-10,5) rectangle (-8, 3);
\draw[line width=2pt,fill=none]  (-8, 5) rectangle (-7, 3);
\draw[line width=2pt,fill=prota2]  (-7, 5) rectangle (-4, 3);
\draw[line width=2pt,fill=none]  (-4, 5) rectangle (-1, 3);
\draw[line width=2pt,fill=prota3] (-1, 5) rectangle ( 0, 3);

\draw[line width=2pt,fill=none] (-10,3) rectangle (-8, 2);
\draw[line width=2pt,fill=none] (-8, 3) rectangle (-7, 2);
\draw[line width=2pt,fill=none] (-7, 3) rectangle (-4, 2);
\draw[line width=2pt,fill=none] (-4, 3) rectangle (-1, 2);
\draw[line width=2pt,fill=none] (-1, 3) rectangle ( 0, 2);

\draw[line width=2pt,fill=prota2] (-10,2) rectangle (-8, -1);
\draw[line width=2pt,fill=none] (-8, 2) rectangle (-7, -1);
\draw[line width=2pt,fill=prota1]  (-7, 2) rectangle (-4, -1);
\draw[line width=2pt,fill=none] (-4, 2) rectangle (-1, -1);
\draw[line width=2pt,fill=prota4] (-1, 2) rectangle ( 0, -1);

\draw[line width=2pt,fill=none] (-10,-1) rectangle (-8, -4);
\draw[line width=2pt,fill=none] (-8, -1) rectangle (-7, -4);
\draw[line width=2pt,fill=none] (-7, -1) rectangle (-4, -4);
\draw[line width=2pt,fill=none] (-4, -1) rectangle (-1, -4);
\draw[line width=2pt,fill=none] (-1, -1) rectangle ( 0, -4);

\draw[line width=2pt,fill=prota3] (-10,-4) rectangle (-8, -5);
\draw[line width=2pt,fill=none]  (-8, -4) rectangle (-7, -5);
\draw[line width=2pt,fill=prota4]  (-7, -4) rectangle (-4, -5);
\draw[line width=2pt,fill=none]  (-4, -4) rectangle (-1, -5);
\draw[line width=2pt,fill=prota1]   (-1, -4) rectangle ( 0, -5);

\draw[ >=latex, ->, line width=3pt ] (1, 0) -- (4,0);

\draw[] (5,-3) rectangle (11,3);

\draw[line width=2pt,fill=prota1] (5,  3) rectangle (7,  1);
\draw[line width=2pt,fill=prota2] (7,  3) rectangle (10, 1);
\draw[line width=2pt,fill=prota3] (10, 3) rectangle (11, 1);

\draw[line width=2pt,fill=prota2] (5,  1) rectangle (7,  -2);
\draw[line width=2pt,fill=prota1] (7,  1) rectangle (10, -2);
\draw[line width=2pt,fill=prota4] (10, 1) rectangle (11, -2);

\draw[line width=2pt,fill=prota3] (5,  -2) rectangle (7,  -3);
\draw[line width=2pt,fill=prota4] (7,  -2) rectangle (10, -3);
\draw[line width=2pt,fill=prota1] (10, -2) rectangle (11, -3);

\end{axis}

\end{tikzpicture}
  \end{minipage}
  \caption{Batch matrix compression scheme for operator basis representations
  relative to non-negligible function indices. Colored tiles represent matrix
  elements which are to be included in the compressed matrix, and white tiles
  represent matrix elements which are to be neglected. Note that these do not
  necessarily correspond to zeros / non-zeros in the original matrix.}
  \label{fig:mat_compress}
\end{figure}
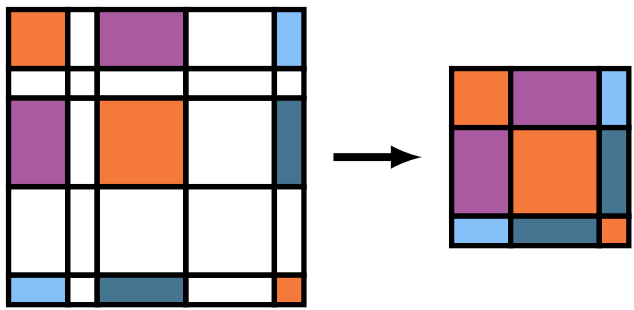

As written, the GEMM and SYR2K given in \cref{eq:x_mat,eq:vxc_batch} are block
sparse level-3 BLAS operations, i.e. BLAS operations involving
matrices which contain many blocks which are numerically zero. To avoid
performing unnecessary FLOPs in the evaluation of these intermediates, it is
possible to store the batch local matrices in
\cref{eq:batch_local_mats,eq:x_mat,eq:vxc_batch} in a compressed format which
stores the blocks corresponding to non-negligible basis shells contiguously and
explicitly removes the zeros from related computation
\cite{stratmann96_achieving}.  A pictorial representation of this matrix
compression for the density matrix is given in \cref{fig:mat_compress}.  We
note for completeness that the forms of \cref{eq:x_mat,eq:vxc_batch} do not
change under this compression, but the sizes of the free indices (as well as
the contracted index in the case of \cref{eq:x_mat}) are reduced.  To avoid a
full decompression of the batched $\mathbf{V}^j$ intermediates,
\cref{eq:vxc_inc} may be implemented by simply incrementing the blocks of the
full dimensional $\mathbf{V}^{xc}$ by the corresponding blocks of
$\mathbf{V}^j$ for each $j$. Note that compression of $\boldsymbol{\Phi}^j$,
$\mathbf{X}^j$, and $\mathbf{Z}^j$ need not be explicit in that they may be 
evaluated directly in compressed form.
 
\ifarxivtypeset
\else
  ~\\
\fi
\subsection{Distributed Parallel Implementation on Clusters of GPU Accelerators}
\label{sec:dist_xc_eval}

In this section, we propose a three-level parallelism scheme for the
distributed evaluation of $\mathbf{V}^{xc}$ and $\mathcal{E}^{xc}$.  A
schematic representation of this procedure is given in \cref{alg:dist_xc_eval}.
For simplicity in the following discussion, we will assume MPI message passing
for distributed computation. Parallelism will be expressed at the following
levels:
\begin{enumerate}
  \item concurrent evaluation of the quadrature batches between independent
  computing ranks,
  \item concurrent evaluation of the quadrature batches assigned to a particular
  computing rank,
  \item and concurrency within the evaluation of a particular quadrature batch
  to evaluate terms such as the atomically-scaled quadrature weights,  batch
  collocation and local density matrices, the level-3 BLAS operations of
  \cref{eq:x_mat,eq:vxc_batch}, etc.
\end{enumerate} 
In the context of the batching scheme discussed in \cref{sec:num_int}, ensuring
proper local sparsity in the batch local $\mathbf{P}^j$ and
$\boldsymbol{\Phi}^j$ typically generates a large number of relatively small
batches which must be evaluated.  As the work required to evaluate a single
$\mathcal{B}_j$ is typically small, distributing its evaluation would  be
be inefficient.  Given that $\mathbf{P}$ and $\mathbf{V}^{xc}$ can be
replicated 
in the memory spaces accessible to each the compute rank, the evaluation of
each quadrature batch requires no communication, Thus the fully distributed
numerical integration of the XC quantities may be performed with only a single
distributed reduction (\texttt{MPI\_Reduce} or \texttt{MPI\_Allreduce}) following
completely independent local computation. We note for posterity that this replication
need not constitute a unique copy of these matrices for each compute rank, only
that these matrices are accessible from each rank, e.g. in the case of partitioned
global address space (PGAS) distributed memory models such as the one provided
by the \texttt{GlobalArrays} library, it would be sufficient to keep a single copy of 
\begin{algorithm2e}[t]
  \caption{Parallelism Scheme for the Evaluation of the XC Potential 
  and XC Energy}
  \label{alg:dist_xc_eval}
  \SetKwInOut{kinput}{Input}
  \SetKwInOut{koutput}{Output}
  \SetKwRepeat{Do}{do}{while}

  \BlankLine
  \BlankLine
  \kinput{ Density matrix $\mathbf{P}$, basis functions $\mathcal{S}$ and atomic centers $\mathcal{A} = \{\mathbf{R}_A\}$. }
  \BlankLine
  \koutput{ XC potential $\mathbf{V}^{xc}$, XC energy $\mathcal{E}^{xc}$. }
  \BlankLine
  \BlankLine
  
  \nl \makebox[10.5cm][l]{
    $\mathcal{B}_{local} \leftarrow$ Form balanced local 
    batches according to \cref{alg:load_balance}.
  } (host) \\
  \BlankLine
  \nl \makebox[10.5cm][l]{
    Perform device allocation.
  } (host/device) \label{alg_ln:device_alloc}\\
  \BlankLine
  \nl \makebox[10.5cm][l]{
    Send constant data (e.g. $\mathbf{P}$, $\mathcal{S}$, and $\mathcal{A}$) to the device.
  } (host/device) \\
  \BlankLine
  \nl \makebox[10.5cm][l]{
    $\mathbf{V}_{local} \leftarrow 0$; $\mathcal{E}_{local}\leftarrow 0$.
  } (device) \\
  \BlankLine
  \Do{$\mathcal{B}_{local} \neq \emptyset$}{
    \nl ~$\mathcal{B}_{device}\leftarrow$ Determine subset of $\mathcal{B}_{local}$ to \\
    \makebox[9.9cm][l]{
      saturate device memory
    } (host) \label{alg_ln:device_batch_subset}\\
    \BlankLine
    \nl \makebox[9.9cm][l]{
      $\mathcal{B}_{local} \leftarrow \mathcal{B}_{local} \setminus \mathcal{B}_{device}$
    } (host) \label{alg_ln:remove_subset}\\
    \BlankLine
    \nl \makebox[9.9cm][l]{
      Pack $\mathcal{B}_{device}$ contiguously on host and send to device.
    } (host/device) \label{alg_ln:pack_and_send_batches}\\
    \BlankLine
    \nl \makebox[9.9cm][l]{
      Update $\mathbf{V}_{local}$ and $\mathcal{E}_{local}$ by 
      $\mathcal{B}_{device}$ according to \cref{alg:process_batch}.
    } (device) \label{alg_ln:process_batch_device}\\
    \BlankLine
  }
  \BlankLine
  \nl \makebox[10.5cm][l]{
    Retrieve $\mathbf{V}_{local}$ and $\mathcal{E}_{local}$ from device 
  } (host/device) \\
  \BlankLine
  \nl \makebox[10.5cm][l]{
    (All) reduce $\mathcal{E}^{xc} \leftarrow \mathcal{E}_{local}$ 
  } (host) \label{alg_ln:e_reduce}\\
  \BlankLine
  \nl \makebox[10.5cm][l]{
    (All) reduce $\mathbf{V}^{xc} \leftarrow \mathbf{V}_{local}$ 
  } (host) \label{alg_ln:v_reduce}\\
  \BlankLine
  \BlankLine
  \Return $(\mathbf{V}^{xc}, \mathcal{E}^{xc})$
  \BlankLine
  
\end{algorithm2e}
these matrices within the memory accessible to a single compute node. However, in
this work, we do not explore the use of PGAS memory models, thus the replication
will be performed at the rank level.

\subsubsection{Distributed Load Balance in the XC Integration}
\label{sec:load_balance}

Despite this embarrassingly parallel integration procedure, care must be taken
to ensure load balance among the independent ranks as the variance in the
computational work required between different batches is often quite large due
to differences in local sparsity and batch sizes.  The simplest choice to
distribute this work would be to distribute the batches at the atomic
quadrature level, i.e. each rank receives the quadrature batches generated from
a particular atomic quadrature. However, this scheme can lead to load imbalance
as the local sparsity of the atoms far from the center of mass can often be
much larger than those which are surrounded by other atoms. In this work, we
choose to distribute the work at the individual batch level by approximating
the FLOPs incurred by each batch,
\begin{equation}
W_j = N_g^j \left( N_A^2 + 9N_b^j + 2(N_b^j)^2 + 3\right) + (N_b^j)^2. \label{eq:batch_work}
\end{equation}
Note that $W_j$ does not represent the true number of FLOPs required to evaluate
intermediates associated with $\mathcal{B}_j$, e.g.  we do not consider FLOP estimates for evaluation of the
exponential in \cref{eq:gau_cart}, nor screening in the evaluation
of the atomic weight scaling, etc. However, $W_j$
has empirically sufficed to produce balanced distributed
computation for all problems considered.  A schematic for the load balance 
scheme used in this work is given in \cref{alg:load_balance}. There are two
important remarks that should be understood from \cref{alg:load_balance}.  The
first is that it requires no communication between independent ranks, i.e.  the
load balance is replicated on each processor. The second is that once the set
of local batches $\mathcal{B}_{local}$ has been determined for each processor,
batches with the same $\mathcal{S}_j$ are merged into a single batch
(\cref{alg_ln:local_merge}). The rationale behind this step is to avoid
polluting the device memory with redundant copies of $\mathbf{P}^j$ and
$\mathbf{V}^j$. 

While \cref{alg:load_balance} could be implemented on the GPU, as has been
discussed in the context of batch generation in related work
\cite{madushanka20_parallel}, we do not explore such implementations in this
work. To improve the performance of the CPU implementation of
\cref{alg:load_balance}, the loop around the atomic quadrature
batches may be parallelized using shared memory parallelism schemes such as
\texttt{OpenMP}. Further, as has been suggested by others
\cite{yasuda08_accelerating}, the cost of grid generation may be amortized in
calculations involving many Fock matrix formations with the same nuclear
geometry by forming it once for the formation of the first Fock matrix and
reusing it for subsequent formations. As will be demonstrated in
\cref{sec:results}, \cref{alg:load_balance} only becomes a computational bottleneck in the
strong scaling limit for medium-to-large molecular systems.

\begin{algorithm2e}[t]
  \caption{Quadrature Batch Load Balance for Distributed XC Integration}
  \label{alg:load_balance}
  \SetKwInOut{kinput}{Input}
  \SetKwInOut{koutput}{Output}
  \DontPrintSemicolon

  \BlankLine
  \BlankLine
  \kinput{ Basis functions $\mathcal{S}$ and atomic centers $\mathcal{A} = \{\mathbf{R}_A\}$. }
  \BlankLine
  \koutput{ Local quadrature batches $\mathcal{B}_{local}$. }
  \BlankLine
  \BlankLine

  \nl $myRank\leftarrow$ Current MPI rank.\\
  \BlankLine
  \nl Compute $\{r_\mu^{cut}\}$ via \cref{eq:rcut} for $\phi_\mu \in \mathcal{S}$. \\
  \BlankLine
  \nl $\mathcal{W}\leftarrow$ Allocate an array of size of the number MPI ranks. \\
  \BlankLine
  \nl $\mathcal{W}\leftarrow 0$; $\mathcal{B}_{local}\leftarrow\emptyset$ \\
  \BlankLine
  \For{ $\mathbf{R}_A \in \mathcal{A}$ } {
    \BlankLine
    \nl $\mathcal{Q}_A\leftarrow$ Form spherical quadrature around $\mathbf{R}_A$.\\
    \BlankLine
    \nl $\mathcal{B}_A\leftarrow$ Generate batches from $\mathcal{Q}_A$.
      \label{alg_ln:atomic_batch_generate}\\
    \BlankLine
    \For{ $\mathcal{B}_j \in \mathcal{B}_A$ } {
      \BlankLine
      \nl $\mathcal{S}_j\leftarrow$ Select from $\mathcal{S}$ the non-negligible basis functions via \cref{alg:cube_sphere_intersection}
      with the cuboid enclosing $\mathcal{B}_j$ and the spheres defined by $\{\mathbf{R}_\mu\}$ and $\{r_\mu^{cut}\}$.
      \label{alg_ln:basis_screen}\\
      \BlankLine
      \nl $W_j\leftarrow$ Compute work estimate for $\mathcal{B}_j$ via \cref{eq:batch_work}.\\
      \BlankLine
      \nl $I\leftarrow$ Find rank with minimum workload from $\mathcal{W}$.\\
      \BlankLine
      \nl $\mathcal{W}_I\leftarrow \mathcal{W}_I + W_j$.\\
      \BlankLine
      \lIf{$I = myRank$}{$\mathcal{B}_{local} \leftarrow \mathcal{B}_{local} \cup \{\mathcal{B}_j\}$.}
    }
  }
  \BlankLine
  \BlankLine
  \nl $\mathcal{B}_{local}\leftarrow$ Merge $\mathcal{B}_j\in\mathcal{B}_{local}$ with the same $\mathcal{S}_j$.
    \label{alg_ln:local_merge}\\
  \BlankLine
  \BlankLine
  \Return $\mathcal{B}_{local}$
\end{algorithm2e}

\subsubsection{Local XC Integration on the GPU}
\label{sec:xc_gpu}

Up to this point, the discussed work distribution scheme has been largely independent
of whether or not the evaluation of local quadrature batches is to be performed on the
host or the device. In this work, we only consider the case where a single MPI rank 
is driving a single device (one-to-one), i.e. we do not consider device affinities of
multiple MPI ranks driving a single device (many-to-one) nor a single MPI rank driving multiple devices
(one-to-many). The method proposed could be extended to one-to-many
device affinities through an additional invocation of \cref{alg:load_balance} to produce balanced
quadrature batches which are to be executed on a particular device. However, in the strong scaling
limit, it would be unlikely that this affinity would be resource efficient due to a
decrease in work assigned to any particular compute rank. 

\paragraph{Architecture of NVIDIA Tesla V100}
\label{sec:v100}

The GPU targeted in this work is the NVIDIA Tesla V100-SXM2 using the CUDA
programming environment.  However, the methodological developments described in
this work may be extended to any GPU device given a software stack which
provides batched BLAS functionality. The V100 is equipped with 16GB
high-bandwidth global memory and 80 streaming multiprocessors (SM). Within the CUDA
model, independent tasks are launched in the form of kernels and concurrency on
the device is expressed in a four level parallelism scheme: 
\begin{itemize}
  \item At the lowest level is the GPU thread which executes instructions
  issued by the SM.

  \item In contrast to CPU architectures, where all threads may execute more
  or less independently, the overhead of instruction issuance is mitigated on GPU
  devices in part by issuing a single instruction to multiple threads which execute
  in lock step. This is known as single-instruction multiple thread (SIMT)
  concurrency, and the collection of threads which execute in this manner is
  known as a \emph{warp} in the CUDA vernacular. On the V100, a warp consists
  of 32 threads. 

  \item Warps are then collected into groups called thread blocks which may
  share data and be mutually synchronized. Thread blocks are typically comprised
  of 256-1024 threads which execute independently at the warp level. 

  \item Thread blocks are further grouped into process grids which are specified
  at the time that the kernel is launched. A kernel has completed once all the
  thread blocks in its specified process grid have finished executing.
\end{itemize}
For a kernel launched with a particular process grid, thread blocks are
scheduled and executed concurrently among the different SM's. Ordering of
kernel execution on CUDA devices is achieved by a software construct known as a
stream: kernels launched on the same stream are guaranteed to be executed in
the order with which they were specified. For kernels which are designed not to
achieve full occupancy within the SM,
it is possible to overlap independent kernel invocations on
separate streams. In this work, however; the kernels developed are designed to
achieve high occupancy within each SM, thus the potential for overlap of
independent kernels is minimal. 
Another consideration one must account for within the SIMT execution model is
the concept of warp divergence, i.e.  kernels which execute different
instructions within a particular warp. Due to the SIMT execution model,
instructions must be executed at the warp level, thus if branch logic causes
the warp to diverge into $N$ unique instructions, the execution time of this
kernel will be roughly the sum of the execution times for the individual
instructions, thus reducing the parallel efficiency of the particular kernel.
Such divergence can lead to significant performance degradation. As such, one
must carefully design GPU kernels such that unique instructions which are
desired to execute concurrently are executed along (or near) warp boundaries to
avoid such degradation.

\paragraph{Data Locality}
\label{sec:data_locality}

The algorithm presented in this work aims to maximize the
potential for concurrency in the evaluation of the local quadrature batches
by minimizing synchronization points, such as data transfers and memory
allocations, which hinder the ability to express concurrency.  As the
computational work required to evaluate any particular quadrature batch is
small, concurrency is achieved by batching the evaluation of the quadrature
batches on the GPU.  This approach has been inspired by GPU accelerated batched
BLAS operations which achieve high throughput by batching
the evaluation of small matrix operations into a single kernel launch
\cite{haidar15_batched,abdelfattah16_performance}. Given that the data
associated with a particular $\mathcal{B}_j$ must reside in device memory for
it to be processed (quadrature points and weights, $\mathcal{S}_j$,
$\boldsymbol{\Phi}^j$, $\mathbf{P}^j$, $\mathbf{Z}^j$, etc.), the approach
taken in this work is to saturate the
device memory with as many quadrature batches as possible as to allow for their
concurrent evaluation.  Note that this approach does not change the amount of
data that must be transferred between host and device throughout the XC
integration, but it does reduce the frequency and improve the performance of
these data transfers by saturating the bandwidth between host and device while
allowing for the expression of more concurrency on the device between data
transfers.  In the case when all of the quadrature batches are unable to
simultaneously occupy the device memory, subsets of the local quadrature
batches which saturate device memory are chosen to be executed concurrently
until all batches have been processed. A depiction of this procedure is given
in
\cref{alg_ln:device_batch_subset,alg_ln:remove_subset,alg_ln:pack_and_send_batches,alg_ln:process_batch_device}.
The performance of these data transfers may be further improved in
\cref{alg_ln:pack_and_send_batches} by packing the batch data contiguously into
page-locked memory (as is produced by \texttt{cudaMallocHost} in the CUDA SDK) on the host.
In addition, rather than perform numerous memory allocations and deallocations
between processing subsets of local quadrature batches, the cost of device
memory allocation may be amortized by preallocating a large fraction of
available device memory at the beginning of the XC integration and manually
managing memory allocation throughout the calculation
(\cref{alg_ln:device_alloc}).  Note that a vast majority of the data associated
with a particular $\mathcal{B}_j$ need not be referenced on the host nor
transferred between host and device. In essence, the only batch specific data
that need be transferred between host and device for a particular
$\mathcal{B}_j$ are its quadrature points and weights, the information
pertaining to the atomic center which generated that batch (for the evaluation of the atomic partition function), and the information
describing $\mathcal{S}_j$. All other data may be allocated and manipulated
directly on the device.

In addition to batch specific data which must reside in device memory, there
are a number of other quantities which are unrelated to a particular batch
that are useful to store in device memory to avoid host-device transfers and to
exploit the high-bandwidth memory which is common on contemporary devices.
These quantities include $\mathbf{P}$, $\mathcal{S}$ and things such as the
atomic positions, inter-nuclear distances, etc. For example, in cases where
$\mathbf{P}$ can reside in memory, the packing of batch local $\mathbf{P}^j$
may be made very efficient by limiting data transfers to be internal to the
device memory (i.e. device memory copies).  In addition, it is also
advantageous to store local contributions to $\mathbf{V}^{xc}$ and
$\mathcal{E}^{xc}$ on the device as to avoid communication of intermediate data
between the evaluation of batch subsets on the device.  We note that even for
the largest problem considered in this work (1231 atoms, $N_b=$ O(10,000)), both $\mathbf{V}^{xc}$ and
$\mathbf{P}$ may reside simultaneously in device memory while leaving enough
additional memory for batch specific data as to allow for enough concurrency to
be resource efficient on the device. 
For hypothetical problems for which this is not possible, the packing of $\mathbf{P}^j$
and the increment of $\mathbf{V}^j$ can be performed on the host at the cost of significant 
performance degradation. We do not explore such implementations here.

\ifarxivtypeset
\else
  ~\\
\fi
\paragraph{Batch Execution of Quadrature Batches on the GPU}
\label{sec:batch_execution}

\begin{algorithm2e}[t]
  \caption{Concurrent Evaluation of Quadrature Batches on a GPU Device}
  \label{alg:process_batch}
  \SetKwInOut{kinput}{Input}
  \SetKwInOut{koutput}{Output}
  \SetKwFor{PFor}{parallel for}{do}{end}
  \DontPrintSemicolon

  \BlankLine
  \BlankLine
  \kinput{ Quadrature batches $\mathcal{B}$, density matrix $\mathbf{P}$, XC potential $\mathbf{V}^{xc}$, and 
  XC energy $\mathcal{E}^{xc}$ all in device memory.}
  \BlankLine
  \koutput{ $\mathbf{V}^{xc}$ and $\mathcal{E}^{xc}$ updated by quadrature contributions from $\mathcal{B}$}
  \BlankLine
  \BlankLine

  \PFor{$\mathcal{B}_j \in \mathcal{B}$}{
    \BlankLine
    \nl Update quadrature weights by atomic partition function.\\
    \BlankLine
    \nl $\mathbf{P}^j\leftarrow$ Compress batch local density matrix from $\mathbf{P}$.\\
    \BlankLine
    \nl $(\boldsymbol{\Phi}^j, \nabla\boldsymbol{\Phi}^j)\leftarrow$ Evaluate compressed batch local collocation matrix and its gradient given $\mathcal{S}_j$.\\
    \BlankLine
  }

  \BlankLine
  \BlankLine
  \nl $\{\mathbf{X}^j\}\leftarrow$ Concurrent evaluation of \cref{eq:x_mat} for all $\boldsymbol{\Phi}^j$ and $\mathbf{P}^j$
    via VB-GEMM. \label{alg_ln:vb-gemm}
  \BlankLine
  \BlankLine

  \PFor{$\mathcal{B}_j \in \mathcal{B}$}{
    \BlankLine
    \nl $(\boldsymbol{\densitySymb}^j,\nabla\boldsymbol{\densitySymb}^j)\leftarrow$ 
      Evaluate $\densitySymb$ and $\nabla\densitySymb$ via \cref{eq:eval_local_den,eq:eval_local_den_grad}.\\
    \BlankLine
    \nl$(\boldsymbol{\varepsilon}^j,\boldsymbol{\varepsilon}_\rho^j,\boldsymbol{\varepsilon}_\gamma^j)\leftarrow$ 
      Evaluate XC functional and its derivatives according to \cref{eq:eval_xc_batch}.\\
    \BlankLine
    \nl Update $\mathcal{E}^{xc}$ according to \cref{eq:exc_batch}.\\
    \BlankLine
    \nl $\mathbf{Z}^j\leftarrow$ \cref{eq:z_mat}. \\
    \BlankLine
  }

  \BlankLine
  \BlankLine
  \nl $\{\mathbf{V}^j\}\leftarrow$ Concurrent evaluation of \cref{eq:vxc_batch} for all $\boldsymbol{\Phi}^j$ and $\mathbf{Z}^j$
    via VB-SYR2K. \label{alg_ln:vb-syr2k}
  \BlankLine
  \BlankLine

  \PFor{$\mathcal{B}_j \in \mathcal{B}$}{
    \BlankLine
    \nl Update $\mathbf{V}^{xc}$ by $\mathbf{V}^j$ via \cref{eq:vxc_inc}.
    \BlankLine
  }
\end{algorithm2e}

Given a set of quadrature batches which saturate device memory,
\cref{alg:process_batch} depicts a general outline of the concurrency pattern
for their simultaneous evaluation on a single device. \Cref{alg:process_batch}
exhibits a number of important features which warrant brief discussion.
The first is the utilization of batched level-3 BLAS primitives for the concurrent
evaluation of \cref{eq:x_mat,eq:vxc_batch} for all batches that reside in 
device memory (\cref{alg_ln:vb-gemm,alg_ln:vb-syr2k}). An important remark
related to this batched BLAS invocation is that the batch local matrices
are often not of uniform dimension for all batches in device memory. As such,
they may not be implemented by uniform batched BLAS implementations, such as those
provided by \texttt{cuBLAS}. In this work, we have used the variable-dimension batched
(or ``vbatched") GEMM (VB-GEMM) and SYR2K (VB-SYR2K) implementations from 
the MAGMA \cite{tdb10,ntd10,tensors} library to perform these batched evaluations.
Another important feature of \cref{alg:process_batch} is that, while the order
of operations within the various \textbf{parallel for} loops are indicative of
the order with which the various tasks are executed at a high level,  each of
these tasks represent individual kernels for which concurrency between the
separate $\mathcal{B}_j$'s occurs at the thread block level. That is to say
that each kernel invocation performs the \textbf{parallel for} loop as a batched 
invocation for each task individually. As has been discussed in similar work \cite{laqua20_highly},
these operations could also be scheduled on different streams to achieve concurrency
in batch execution. 
We do not explore such implementations in this work.
Finally, much like the batched BLAS invocations which are designed to express
concurrency both within a matrix operation and between matrix operations
themselves, each kernel invocation for the XC specific tasks in
\cref{alg:process_batch} is designed to express concurrency within each task as
well. 
Each batch-local task is designed to occupy a subset
of the process grid while evaluation of each batch local task is performed
independently on separate subsets within the same kernel launch. In practice,
this may implemented using multi-dimensional kernel launches within the CUDA
framework.


While GPU accelerated BLAS functionality may be provided by optimized
third-party libraries, as of this work there does not exist standard GPU
implementations of the remainder of the operations required for the XC
integration. As such, they must be implemented and optimized by hand. 
The details of such implementations are outside the scope of this work as they
are largely dependent on the data structures used in a particular software.
However, there a few important details related to the algorithmic choices used
in this work which warrant brief discussion.
%
%
In the context of the evaluation of $\boldsymbol{\Phi}^j$ on the device, we
adopt a simple strategy which assigns the evaluation of a single contracted
basis shell  at a particular point to a single thread, i.e. we do no express
concurrency in the evaluation of the exponential factors of the primitive
Gaussians. Care is taken in the implementation presented in this work to
minimize the chance of warp divergence by assigning evaluations of the same
basis shell at various quadrature points to the same warp (i.e. to minimize the
frequency of divergence in the sum of \cref{eq:gau_cart} with functions of
differing $n_\xi^\mu$). We will demonstrate the efficacy of this simple
strategy in \cref{sec:results}. 

A major difference in the work presented here relative to existing methods for GPU
XC integration \cite{yasuda08_accelerating,madushanka20_parallel} is the strategy for the
evaluation of $\boldsymbol{\epsilon}^j$ and its functional derivatives on the
device. On the CPU, there are several standard libraries, such as \texttt{LIBXC}
\cite{lehtola18_recent} and \texttt{XCFun} \cite{ekstrom_ulf_2020}, which
implement a vast number of XC functionals which are commonly used in KS-DFT
calculations. Some work \cite{madushanka20_parallel} has been dedicated to
porting all, or portions of these libraries to the GPU, including an initial
implementation of porting \texttt{LIBXC} to CUDA in the development version of
the library itself. However, there does not exist a mature, high-performance 
GPU interface for these libraries at this time. To ensure the highest
performance possible, the approach taken in this work has been to develop 
an open-source library, \texttt{ExchCXX} \cite{exchcxx}, which provides the
necessary functionality. \texttt{ExchCXX} is a modern C++ library which
implements a commonly used subset of XC functionals for evaluation on the host
or device though a simple, common API.  We note that the numerical expressions
for the XC functionals implemented in \texttt{ExchCXX} have been taken directly
from \texttt{LIBXC} and have been demonstrated to produce numerically
indistinguishable results. 

We note for posterity that, in previous work \cite{yasuda08_accelerating}, the
use of single precision and mixed precision arithmetic has been shown to
further improve the performance of GPU accelerated XC integration. However,
as the performance gap between single and double precision arithmetic on GPU
hardware has been closing in recent years \cite{cook12_cuda}, all
calculations performed in this work use strictly double-precision arithmetic.

\section{Results}
\label{sec:results}

\begin{table}[t]
\centering
\caption{Molecule Sizes and Basis Dimensions}
\label{tbl:molecules}
\begin{tabular}{llll}
\hline
Molecule     & $N_A$ & $N_b$ / 6-31G(d) & $N_b$ / cc-pVDZ \\
\hline
Taxol        & 110  & 1013  & 1099   \\
Valinomycin  & 168  & 1350  & 1542   \\
Olestra      & 453  & 3181  & 3840   \\
Ubiquitin    & 1231 & 10292 & 11577  \\
\hline
\end{tabular}
\end{table}

\begin{table}[t]
\centering
\caption{Atomic Quadrature Sizes}
\label{tbl:grids}
\begin{tabular}{llll}
\hline
Grid         & $N_{ang}$ & $N_{rad}$ & $N^A_{g}$ \\
\hline
FG           &  302      & 75        & 22650  \\
UFG          &  590      & 99        & 58410  \\
SFG          &  974      & 175       & 170450 \\
\hline
\end{tabular}
\end{table}

In essence, the method proposed and implemented in this work (\cref{alg:dist_xc_eval}) is composed of three computationally
dominant phases:
\begin{enumerate}
  \item a load balancing phase which is replicated on all MPI ranks (\cref{alg:load_balance}),
  \item a local integration phase which is executed on the device (\cref{alg:process_batch}),
  \item and a reduction phase which combines the locally computed XC quantities in distributed
  memory to produce the final integration results.
\end{enumerate}
In this section, we examine various performance characteristics of these phases
as implemented in the open-source NWChemEx software package \cite{nwchemex}.  In
addition, we compare the performance and scaling of this implementation to that
of an analogous scalable CPU implementation in the open-source NWChem software
package \cite{apra20_nwchem}.  We have chosen to examine the performance of the
purposed method as applied to 4 molecules: Taxol, Valinomycin, Olestra, and
Ubiquitin; and 2 basis sets: 6-31G(d)
\cite{ditchfield1971a,francl1982a,gordon1982a,hariharan1973a,hehre1972a} and
cc-pVDZ \cite{dunning1989a,woon1993a}, to provide a performance
characterization for systems with a wide range of size, spacial extent and
basis dimension. The geometries and references for this structures are included 
in the Suppplemental Information. All calculations were
performed using the PBE GGA XC functional \cite{perdew96_generalized}.
Calculations involving the 6-31G(d) basis set were performed using Cartesian
Gaussian functions, while those involving cc-pVDZ were performed using
spherical Gaussian functions.  A list of data relevant to the performance of
calculations involving these systems can be found in \cref{tbl:molecules}.  In
addition, we have examined the use of 3 commonly encountered atomic quadrature
sizes: the fine (FG), ultra-fine (UFG) and super-fine (SFG) grids, as described
in \cref{tbl:grids}.

All calculations have been performed on the Summit supercomputer at the Oak
Ridge Leadership Computing Facility (OLCF). Each Summit node consists of 2 IBM
POWER9 CPUs (2x21 @ 3.8 GHz) and 6 NVIDIA Tesla V100 GPUs. To enable a fair
comparison between NWChem and NWChemEx, each Summit node has been subdivided
into 6 equally sized "resource sets" consisting of 7 CPU cores and 1 GPU. For
calculations involving NWChemEx, concurrency in the CPU execution will be
performed in shared memory to adhere to the one-to-one CPU-to-GPU affinity
previously discussed, i.e.  1 MPI rank with 7 shared memory threads driving a
single GPU.  Note that CPU parallelism is only utilized in the generation of
the local quadrature batches as discussed in \cref{sec:load_balance}, and the
launching of kernels to execute \cref{alg:process_batch} on the GPU is
performed in serial.

Calculations involving NWChem were performed using a locally modified copy of
release version 7.0.0. Code modifications were limited to ensuring that the
radial scaling factors of the MK radial quadrature produced identical atomic
quadratures to those in NWChemEx.  Further, NWChem DFT calculations were performed
with grid pruning disabled and using the SSF atomic partitioning scheme. 
Note that while the quadratures are
identical between the two codes, NWChem exhibits a number of algorithmic
differences with those presented in this work. These include additional density
and weight screening techniques within each quadrature batch. However, these
steps only improve the observed performance in NWChem, thus do not detract from
the performance comparisons made in this work.  To ensure that we are
comparing with consistent, replicatable performance in NWChem, all
calculations have been performed using converged density matrices. Each resource set will
consist of 7 MPI ranks for calculations involving NWChem as, with the exception
of the atomic weight scaling, its implementation of the XC integration does not
exploit shared memory parallelism. Further, we note that the use of the
\texttt{GlobalArrays} library \cite{nieplocha06_advances,krishnan12_global} in
NWChem yields that one MPI rank per physical node will be used as a progress
rank for remote memory access rather than performing computation related to the
XC integration.

Both NWChem and NWChemEx were compiled using the GNU 8.1.0 compiler suite
(\texttt{gcc}, \texttt{g++}, \texttt{gfortran}) to compile host code using high
levels of compiler optimization (\texttt{-O3 -mcpu=native -mtune=native
-ffast-math}). The device code in NWChemEx was compiled using the NVIDIA CUDA
compiler (\texttt{nvcc}) as provided in the CUDA SDK (version 10.1.105). Analogous
optimization flags (\texttt{-O3 {-}{-}use-fast-math}) as well as architecture
specific flags to generate optimized binaries for CUDA compute capability 7.0
(\texttt{-gencode sm\_70,compute\_70}) were used in the compilation of device
code. NWChem was linked to the serial version of the IBM
Engineering Scientific Software Library (ESSL version 6.1.0) for POWER9 optimized
BLAS functionality. GPU accelerated batched BLAS was provided by the MAGMA library
(version 2.5.1) while non-batched BLAS for operations such as dot products 
was provided by the \texttt{cuBLAS} library from the NVIDIA CUDA SDK.

\subsection{Integration Performance on GPU Devices}


\begin{table}
\small
\centering
\caption{Aggregate wall times for computationally intensive operations of XC integration for the various problems 
considered. All times are given in seconds and $N_{sat}$ is the number of times the device memory was 
saturated in \cref{alg:dist_xc_eval} to complete the integration.}
\label{tbl:overall_component}
\begin{tabular}{lllllllllll}
\hline
Molecule     & Basis     & Grid & $N_{sat}$ & Load Balance  & (\%)     & Local Integration  & (\%)     & Other  & (\%)    & Total \\
\hline
Taxol        & 6-31G(d)  & FG   & 1         & 0.073   & (17.49)  & 0.310   & (73.99)  & 0.036  & (8.52)  &  0.419 \\
             &           & UFG  & 2         & 0.145   & (15.50)  & 0.746   & (79.59)  & 0.046  & (4.91)  &  0.937 \\
             &           & SFG  & 3         & 0.252   & (15.76)  & 1.30    & (80.84)  & 0.055  & (3.41)  &  1.60  \\
             & cc-pVDZ   & FG   & 1         & 0.075   & (14.62)  & 0.399   & (77.59)  & 0.040  & (7.79)  &  0.514 \\
             &           & UFG  & 2         & 0.153   & (13.70)  & 0.918   & (82.12)  & 0.047  & (4.18)  &  1.12  \\
             &           & SFG  & 3         & 0.268   & (13.26)  & 1.68    & (83.24)  & 0.071  & (3.50)  &  2.02  \\
\hline
Valinomycin  & 6-31G(d)  & FG   & 1         & 0.128   & (14.74)  & 0.685   & (79.14)  & 0.053  & (6.12)  &  0.865 \\
             &           & UFG  & 3         & 0.259   & (15.79)  & 1.33    & (80.95)  & 0.054  & (3.26)  &  1.64  \\
             &           & SFG  & 5         & 0.446   & (14.98)  & 2.45    & (82.21)  & 0.084  & (2.81)  &  2.98  \\
             & cc-pVDZ   & FG   & 2         & 0.136   & (12.17)  & 0.916   & (82.27)  & 0.062  & (5.55)  &  1.11  \\
             &           & UFG  & 3         & 0.274   & (11.99)  & 1.96    & (85.74)  & 0.052  & (2.27)  &  2.29  \\ 
             &           & SFG  & 6         & 0.474   & (11.09)  & 3.70    & (86.61)  & 0.098  & (2.30)  &  4.27  \\
\hline
Olestra      & 6-31G(d)  & FG   & 2         & 0.433   & (23.60)  & 1.20    & (65.45)  & 0.201  & (10.95) & 1.84   \\
             &           & UFG  & 5         & 0.872   & (23.48)  & 2.65    & (71.39)  & 0.191  & (5.13)  & 3.72   \\
             &           & SFG  & 9         & 1.49    & (21.79)  & 5.14    & (75.13)  & 0.211  & (3.08)  & 6.84   \\
             & cc-pVDZ   & FG   & 3         & 0.481   & (19.87)  & 1.68    & (69.48)  & 0.258  & (10.66) & 2.42   \\
             &           & UFG  & 6         & 0.953   & (19.59)  & 3.63    & (74.57)  & 0.284  & (5.83)  & 4.87   \\
             &           & SFG  & 11        & 1.63    & (18.54)  & 6.92    & (78.53)  & 0.259  & (2.94)  & 8.82   \\
\hline
Ubiquitin    & 6-31G(d)  & FG   & 22        & 3.12    & (10.94)  & 22.5    & (78.90)  & 2.89   & (10.15)&  28.5   \\
             &           & UFG  & 45        & 6.01    & (10.84)  & 47.5    & (85.70)  & 1.92   & (3.46) &  55.4   \\
             &           & SFG  & 84        & 10.2    & (9.94)   & 90.2    & (87.82)  & 2.30   & (2.24) &  103    \\
             & cc-pVDZ   & FG   & 30        & 3.44    & (7.83)   & 38.2    & (86.96)  & 2.29   & (5.21) &  43.9   \\
             &           & UFG  & 61        & 6.64    & (7.50)   & 79.6    & (89.80)  & 2.40   & (2.71) &  88.6   \\
             &           & SFG  & 111       & 11.2    & (7.04)   & 145     & (90.90)  & 3.30   & (2.07) &  160    \\
\hline
\end{tabular}
\end{table}

First, we examine the performance characteristics of
\cref{alg:dist_xc_eval} on a single
Summit node.  This treatment allows us to examine the effects of molecule size,
basis dimension and quadrature size on overall GPU performance separately from
scaling in a distributed setting. Strong scaling of the purposed method, as
well as its comparison to NWChem will be presented in the following subsection.
An overall component analysis of the timings on a single Summit node is given
in \cref{tbl:overall_component}. The wall times presented in
\cref{tbl:overall_component} are aggregated over the entire XC integration,
i.e. for the local integration, the times presented are representative of the
sum of all invocations which saturate device memory ($N_{sat}$).  Further, we
note that these times also include the contiguous host packing and host-device
transfer of batch data (i.e. all operations contained in the loop over
quadrature batches in \cref{alg:dist_xc_eval}).  In addition, the times
presented for load balancing include all operations in \cref{alg:load_balance},
i.e. batch generation and the course-grained screening of basis shells at the
batch level. As these calculations were performed within a single Summit node,
the reduction phase is not explicitly considered in
\cref{tbl:overall_component}, but its contributions are included in the times
labeled ``Other".  As expected, although \cref{alg:load_balance} is executed
on the host in this work, the dominant computational phase for these
calculations was the local integration. Further, we note that the
overall cost of \cref{alg:load_balance} for a particular molecule / grid pair
is largely independent on basis size but scales linearly with respect to grid
size for a particular molecule / basis pair. The result of this is that the
relative cost of load balancing is reduced as basis size increases. However,
while this cost is not dominant at low processor counts, it will be demonstrated
to be dominant in the strong scaling limit in the following subsection.

\begin{figure}
  \centering
  \figname{device-component}
  \begin{tikzpicture}
      
\begin{axis}[
  ybar,
  bar width=6pt,
  enlarge x limits=0.2,
  legend style={font=\tiny,at={(0.5,1.1)},anchor=south},
  legend cell align={left},
  legend columns=5,
  width=\textwidth,
  height=0.35\textwidth,
  ticklabel style = {font=\small},
  label style = {font=\small},
  ytick style={draw=none},
  xtick style={draw=none},
  xtick=data,
  xticklabels={Taxol, Valinomycin, Olestra, Ubiquitin},
  ytick={10,20,30,40},
  cycle list={%
    {draw=black,fill=black!10},%
    {draw=black,fill=black!30},%
    {draw=black,fill=black!50},%
    {draw=black,fill=black},%
    {draw=black,pattern=grid},%
    {draw=black,pattern=north east lines},%
    {draw=black,pattern=north west lines},%
    {draw=black,pattern=crosshatch dots},%
    {draw=black,pattern=crosshatch},%
    {draw=black,fill=black!30,postaction={pattern=north east lines}},%
    {draw=black,fill=black!30,postaction={pattern=north west lines}},%
    {draw=black,fill=black!30,postaction={pattern=grid}},%
    {draw=black,fill=black!30,postaction={pattern=crosshatch}}%
  }, 
  ymin=0,
  ylabel={Wall Time \% }
]

\addlegendentry{SSF};
\addplot+ [area legend] table [
  x=Idx, 
  y expr=100*\thisrow{Weights}/\thisrow{Total},
  col sep=comma
] {fig-src/data/try3/device-component.dat};
\addlegendentry{Form $\boldsymbol{\Phi}^j$};
\addplot+ [area legend] table [
  x=Idx, 
  y expr=100*\thisrow{Collocation}/\thisrow{Total},
  col sep=comma
] {fig-src/data/try3/device-component.dat};
\addlegendentry{VB-GEMM};
\addplot+ [area legend] table [
  x=Idx, 
  y expr=100*\thisrow{VB-GEMM}/\thisrow{Total},
  col sep=comma
] {fig-src/data/try3/device-component.dat};
\addlegendentry{VB-SYR2K};
\addplot+ [area legend] table [
  x=Idx, 
  y expr=100*\thisrow{VB-SYR2K}/\thisrow{Total},
  col sep=comma
] {fig-src/data/try3/device-component.dat};
\addlegendentry{Form $\mathbf{Z}^j$};
\addplot+ [area legend] table [
  x=Idx, 
  y expr=100*\thisrow{Zmat}/\thisrow{Total},
  col sep=comma
] {fig-src/data/try3/device-component.dat};
\addlegendentry{Compute $\densitySymb$ / $\nabla\densitySymb$};
\addplot+ [area legend] table [
  x=Idx, 
  y expr=100*\thisrow{EvalVVars}/\thisrow{Total},
  col sep=comma
] {fig-src/data/try3/device-component.dat};

\addlegendentry{Other};
\addplot+ [area legend] table [
  x=Idx, 
  y expr=100*\thisrow{Other}/\thisrow{Total},
  col sep=comma
] {fig-src/data/try3/device-component.dat};

\addlegendentry{D2H};
\addplot+ [area legend] table [
  x=Idx, 
  y expr=100*\thisrow{D2H}/\thisrow{Total},
  col sep=comma
] {fig-src/data/try3/device-component.dat};

\addlegendentry{H2D};
\addplot+ [area legend] table [
  x=Idx, 
  y expr=100*\thisrow{H2D}/\thisrow{Total},
  col sep=comma
] {fig-src/data/try3/device-component.dat};

\addlegendentry{D2D};
\addplot+ [area legend] table [
  x=Idx, 
  y expr=100*\thisrow{D2D}/\thisrow{Total},
  col sep=comma
] {fig-src/data/try3/device-component.dat};

\end{axis}
\end{tikzpicture}
  \caption{Wall time percentages for various operations in
  the XC integration involving the GPU. Includes host-to-device (H2D),
  device-to-host (D2H) and device-to-device (D2D) transfers.}
  \label{fig:device_component}
\end{figure}
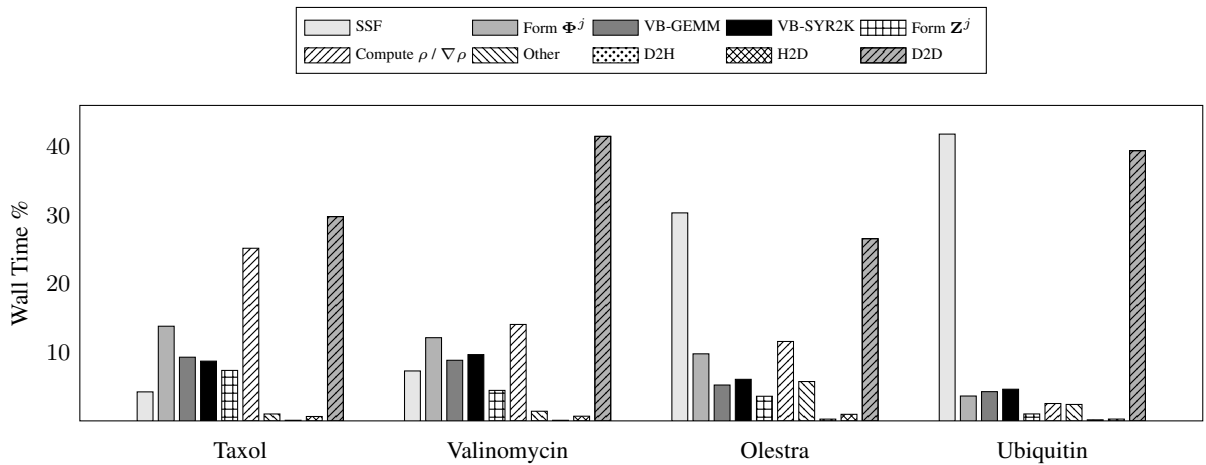

In this work, we focused on two algorithmic motifs that are important for 
the XC integration on the GPU:
\begin{enumerate}
  \item optimizing data locality to minimize the overhead of low-bandwidth
  data transfers between host and device and to maximize the potential
  to express concurrency without synchronization,
  \item and batching together the evaluation of small tasks on the device 
  through the use of kernels which express concurrency both within a quadrature
  batch and between batches to improve throughput on the device.
\end{enumerate}
To demonstrate the efficacy of these motifs, we examine the relative
costs of the various compute and memory intensive operations incurred by the various
kernels during the local integration on the device.  Due to the fact that GPU
computation is generally asynchronous with respect to host computation, care
must be taken in accruing accurate performance data relating to individual
kernels as to not impede computational progress on the device. For this
purpose, we have utilized the NVIDIA profiler \texttt{nvprof} to obtain kernel
level performance metrics.  A summary of the overall time spent on various
 operations involving the GPU for the UFG basis and 6-31G(d)
basis set is provided in \cref{fig:device_component}.

There are a number of important features exemplified in the results presented
in \cref{fig:device_component}. The first is that saturating the device
memory to ensure data locality all but removes the cost of host-to-device (H2D)
and device-to-host (D2H) data transfers, yielding < 1\% of the overall 
computational cost combined for all problems considered. 
For the smaller test cases (Taxol and Valinomycin), the GPU implementation is
dominated by the evaluation of $\densitySymb$ / $\nabla\densitySymb$ and
device-to-device (D2D) memory transfers. For the larger test cases (Olestra and
Ubiquitin), the integration is dominated by the evaluation of the SSF atomic
partition weights and D2D memory transfers. We note for clarity that the D2D
transfers are intra-GPU device memory copies, not inter-GPU communication. The
times for the evaluation of the XC functional on the device are not explicitly
shown in \cref{fig:device_component} as they are negligibly small. They are
however included in the ``Other" timing accumulations.

A somewhat unexpected result is the dominant cost posed by intra-GPU D2D
transfers for all problems considered. The D2D timings including the packing of
\cref{eq:local_density_def}, the incrementing of \cref{eq:vxc_inc}, and various
other small D2D transfers such as those involving storage of the basis
functions, etc. This result is unexpected due to the high-bandwidth of memory
transfers within device memory. To further examine the details of this
unexpected dominant cost, \cref{fig:d2d_throughput} shows the achieved memory
read and write throughputs for the intra-GPU data transfers incurred by the
batch kernels which implement \cref{eq:local_density_def,eq:vxc_inc}. These
achieved throughputs are compared to the peak bandwidth of DDR4 (CPU) memory:
50 GB/s. For these kernels, we are able to achieve a memory throughput of O(100
GB/s) for data writes and between 50-70 GB/s for data reads, with the
throughput for data reads decreasing with increasing system size. This decrease
in data read throughout with system size is likely due to memory bank conflicts
arising from multiple GPU threads accessing the same memory address
simultaneously. Although these kernels are not able to achieve memory
throughput reflective of peak device memory bandwidth (900 GB/s) due to their
access of non-coalesced, non-contiguous memory, they far exceed the throughput
which would be achievable in CPU memory. Further, as the memory footprint of
these packed matrices are among the largest in the purposed method, exploiting
intra-GPU memory transfers avoids additional H2D and D2H transfers which would
pose nontrivial costs due to their low bandwidth.

To demonstrate the efficacy of the batched kernels proposed in this work,
\cref{fig:sm-efficiency,fig:warp-execution} illustrate the capability of these
kernels to efficiently exploit the resources of the device. These figures
present the efficiency of the batched kernels in two regimes.  The SM
efficiency \cref{fig:sm-efficiency} illustrates the efficiency of the kernels
at the SM level by calculating the percentage of time each SM has at least one
active warp executing instructions. The warp execution efficiency
\cref{fig:warp-execution} illustrates their efficiency at the warp level by
calculating the percentage of active threads within each warp in the issuance of
any particular instruction in the kernel execution.  Deviations from 100\% in
the SM efficiency indicate that the SM is sitting idle due to some sort of
contention, e.g. warp divergence, while deviations in the warp execution
efficiency indicate that some warps have diverged such that the SM is only
able to execute instructions to some subset of the threads within these
diverged warps, reducing overall parallel efficiency. These performance
measurements were obtained by the \texttt{nvprof} profiler metrics
\texttt{sm\_efficiency} and \texttt{warp\_execution\_efficiency}, respectively.
As we can see, both the MAGMA provided batched BLAS and the hand optimized XC
integration kernels developed for this work are able to achieve high SM
efficiency, i.e. the SM is occupied and issuing instructions a high percentage
of the time. With the exception of the SSF weights kernel, each of the batched
kernels also exhibits an excellent warp execution efficiency ( > 90\% ), which
means that there are not typically a large number of warp divergences in the
execution of these kernels. The relatively low (60\%-70\%) warp execution
efficiency of the SSF kernels is due to the screening of weight partitions by
the SSF scheme, i.e.  adjacent quadrature points often follow different branch
logic in the screening procedure. Note that the high SM and warp execution
efficiencies for the kernel responsible for the batched evaluation of
$\boldsymbol{\Phi}^j$ by the simple method proposed in this work, combined with
its relatively low cost percentage ( > 20\% ) for all problems considered,
indicate that further optimization of this kernel by more advanced techniques
would likely not yield a large impact on overall wall time.


\begin{figure}
\centering
\figname{d2d-throughput}
\begin{tikzpicture}

\begin{axis}[
  ybar,
  bar width=4pt,
  enlarge x limits=0.15,
  legend style={font=\tiny,at={(0.5,1.1)},anchor=south},
  legend cell align={left},
  legend columns=5,
  width=0.5\textwidth,
  ticklabel style = {font=\small},
  label style = {font=\small},
  ytick style={draw=none},
  xtick style={draw=none},
  xtick=data,
  xticklabels={Taxol, Valinomycin, Olestra, Ubiquitin},
  ytick={50,100,150},
  ymin=0,
  ylabel={Memory Throughput / GB/s},
  cycle list={%
    {draw=black,fill=black!10},%
    {draw=black,fill=black!30},%
    {draw=black,fill=black!50},%
    {draw=black,fill=black},%
    {draw=black,pattern=grid},%
    {draw=black,pattern=north east lines},%
    {draw=black,pattern=north west lines},%
    {draw=black,pattern=crosshatch dots},%
    {draw=black,pattern=crosshatch},%
    {draw=black,fill=black!30,postaction={pattern=north east lines}},%
    {draw=black,fill=black!30,postaction={pattern=north west lines}},%
    {draw=black,fill=black!30,postaction={pattern=grid}},%
    {draw=black,fill=black!30,postaction={pattern=crosshatch}}%
  }, 
]

\addlegendentry{$\mathbf{P}$ Read};
\addplot+ [area legend] table [
  x=Idx, 
  y=DR,
  col sep=comma
] {fig-src/data/try3/d2d-throughput.dat};
\addlegendentry{$\mathbf{P}^j$ Write};
\addplot+ [area legend]table [
  x=Idx, 
  y=DW,
  col sep=comma
] {fig-src/data/try3/d2d-throughput.dat};

\addlegendentry{$\mathbf{V}^j$ Read};
\addplot+ [area legend]table [
  x=Idx, 
  y=VR,
  col sep=comma
] {fig-src/data/try3/d2d-throughput.dat};
\addlegendentry{$\mathbf{V}^{xc}$ Write};
\addplot+ [area legend]table [
  x=Idx, 
  y=VW,
  col sep=comma
] {fig-src/data/try3/d2d-throughput.dat};

\draw[black, dashed] (-0.5,50) -- (3.5,50) node [pos=1/2,above] {\tiny DDR4} node [pos=1/2,below] {\tiny 50 GB/s};

\end{axis}
\end{tikzpicture}
\caption{Achieved memory throughput for dominant D2D data transfers in the XC
integration compared to the peak DDR4 bandwidth in host memory.}
\label{fig:d2d_throughput}
\end{figure}
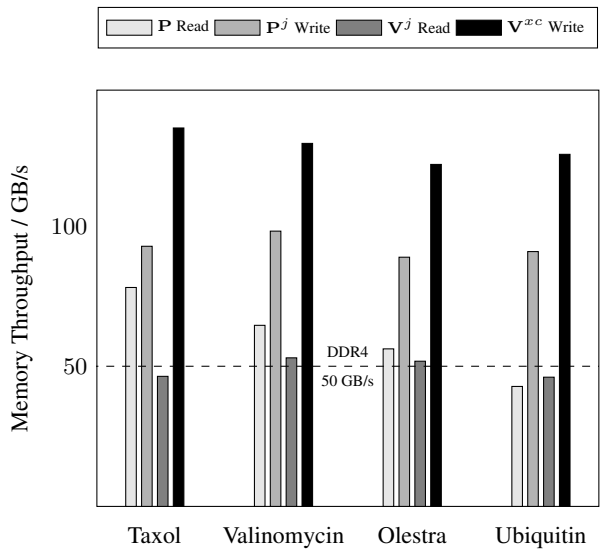

\begin{figure}
\centering
\begin{minipage}{0.49\textwidth}
  \centering
  \figname{sm-efficiency}
  \begin{tikzpicture}

\begin{axis}[
  ybar,
  bar width=4pt,
  enlarge x limits=0.2,
  legend style={font=\tiny,at={(0.5,1.1)},anchor=south},
  legend cell align={left},
  legend columns=5,
  width=\textwidth,
  ticklabel style = {font=\small},
  label style = {font=\small},
  ytick style={draw=none},
  xtick style={draw=none},
  xtick=data,
  xticklabels={Taxol, Valinomycin, Olestra, Ubiquitin},
  ymin=0,
  ylabel={SM Efficiency / \%},
  cycle list={%
    {draw=black,fill=black!10},%
    {draw=black,fill=black!30},%
    {draw=black,fill=black!50},%
    {draw=black,fill=black},%
    {draw=black,pattern=grid},%
    {draw=black,pattern=north east lines},%
    {draw=black,pattern=north west lines},%
    {draw=black,pattern=crosshatch dots},%
    {draw=black,pattern=crosshatch},%
    {draw=black,fill=black!30,postaction={pattern=north east lines}},%
    {draw=black,fill=black!30,postaction={pattern=north west lines}},%
    {draw=black,fill=black!30,postaction={pattern=grid}},%
    {draw=black,fill=black!30,postaction={pattern=crosshatch}}%
  }, 
]

\addlegendentry{VB-GEMM};
\addplot+ [area legend] table [
  x=Idx, 
  y expr=\thisrow{VB-GEMM} * 100,
  col sep=comma
] {fig-src/data/try3/sm-efficiency.dat};
\addlegendentry{VB-SYR2K};
\addplot+ [area legend] table [
  x=Idx, 
  y expr=\thisrow{VB-SYR2K} * 100,
  col sep=comma
] {fig-src/data/try3/sm-efficiency.dat};
\addlegendentry{$\boldsymbol{\Phi}^j$};
\addplot+ [area legend] table [
  x=Idx, 
  y expr=\thisrow{Collocation} * 100,
  col sep=comma
] {fig-src/data/try3/sm-efficiency.dat};
\addlegendentry{$\boldsymbol{\varepsilon}^j$,$\boldsymbol{\varepsilon}^j_\rho$, $\boldsymbol{\varepsilon}_\gamma^j$};
\addplot+ [area legend] table [
  x=Idx, 
  y expr=\thisrow{XCFunc} * 100,
  col sep=comma
] {fig-src/data/try3/sm-efficiency.dat};
\addlegendentry{$\densitySymb$,$\nabla\densitySymb$};
\addplot+ [area legend] table [
  x=Idx, 
  y expr=\thisrow{EvalVVars} * 100,
  col sep=comma
] {fig-src/data/try3/sm-efficiency.dat};
\addlegendentry{SSF};
\addplot+ [area legend] table [
  x=Idx, 
  y expr=\thisrow{Weights} * 100,
  col sep=comma
] {fig-src/data/try3/sm-efficiency.dat};

\end{axis}
\end{tikzpicture}
  \caption{Achieved SM efficiency for batched kernels in the XC integration.}
  \label{fig:sm-efficiency}
\end{minipage}~\hfill
\begin{minipage}{0.49\textwidth}
  \centering
  \figname{warp-execution}
  \begin{tikzpicture}

\begin{axis}[
  ybar,
  bar width=4pt,
  enlarge x limits=0.2,
  legend style={font=\tiny,at={(0.5,1.1)},anchor=south},
  legend cell align={left},
  legend columns=5,
  width=\textwidth,
  ticklabel style = {font=\small},
  label style = {font=\small},
  ytick style={draw=none},
  xtick style={draw=none},
  xtick=data,
  xticklabels={Taxol, Valinomycin, Olestra, Ubiquitin},
  ymin=0,
  ylabel={Warp Execution Efficiency / \%},
  cycle list={%
    {draw=black,fill=black!10},%
    {draw=black,fill=black!30},%
    {draw=black,fill=black!50},%
    {draw=black,fill=black},%
    {draw=black,pattern=grid},%
    {draw=black,pattern=north east lines},%
    {draw=black,pattern=north west lines},%
    {draw=black,pattern=crosshatch dots},%
    {draw=black,pattern=crosshatch},%
    {draw=black,fill=black!30,postaction={pattern=north east lines}},%
    {draw=black,fill=black!30,postaction={pattern=north west lines}},%
    {draw=black,fill=black!30,postaction={pattern=grid}},%
    {draw=black,fill=black!30,postaction={pattern=crosshatch}}%
  }
]

\addlegendentry{VB-GEMM};
\addplot+ [area legend] table [
  x=Idx, 
  y expr=\thisrow{VB-GEMM} * 100,
  col sep=comma
] {fig-src/data/try3/warp-execution.dat};
\addlegendentry{VB-SYR2K};
\addplot+ [area legend] table [
  x=Idx, 
  y expr=\thisrow{VB-SYR2K} * 100,
  col sep=comma
] {fig-src/data/try3/warp-execution.dat};
\addlegendentry{$\boldsymbol{\Phi}^j$};
\addplot+ [area legend] table [
  x=Idx, 
  y expr=\thisrow{Collocation} * 100,
  col sep=comma
] {fig-src/data/try3/warp-execution.dat};
\addlegendentry{$\boldsymbol{\varepsilon}^j$,$\boldsymbol{\varepsilon}^j_\rho$, $\boldsymbol{\varepsilon}_\gamma^j$};
\addplot+ [area legend] table [
  x=Idx, 
  y expr=\thisrow{XCFunc} * 100,
  col sep=comma
] {fig-src/data/try3/warp-execution.dat};
\addlegendentry{$\densitySymb$,$\nabla\densitySymb$};
\addplot+ [area legend] table [
  x=Idx, 
  y expr=\thisrow{EvalVVars} * 100,
  col sep=comma
] {fig-src/data/try3/warp-execution.dat};
\addlegendentry{SSF};
\addplot+ [area legend] table [
  x=Idx, 
  y expr=\thisrow{Weights} * 100,
  col sep=comma
] {fig-src/data/try3/warp-execution.dat};

\end{axis}
\end{tikzpicture}
  \caption{Achieved warp execution efficiency for batched kernels in the XC integration.}
  \label{fig:warp-execution}
\end{minipage}
\end{figure}

\subsection{Strong Scaling}
\label{sec:strong-scaling}

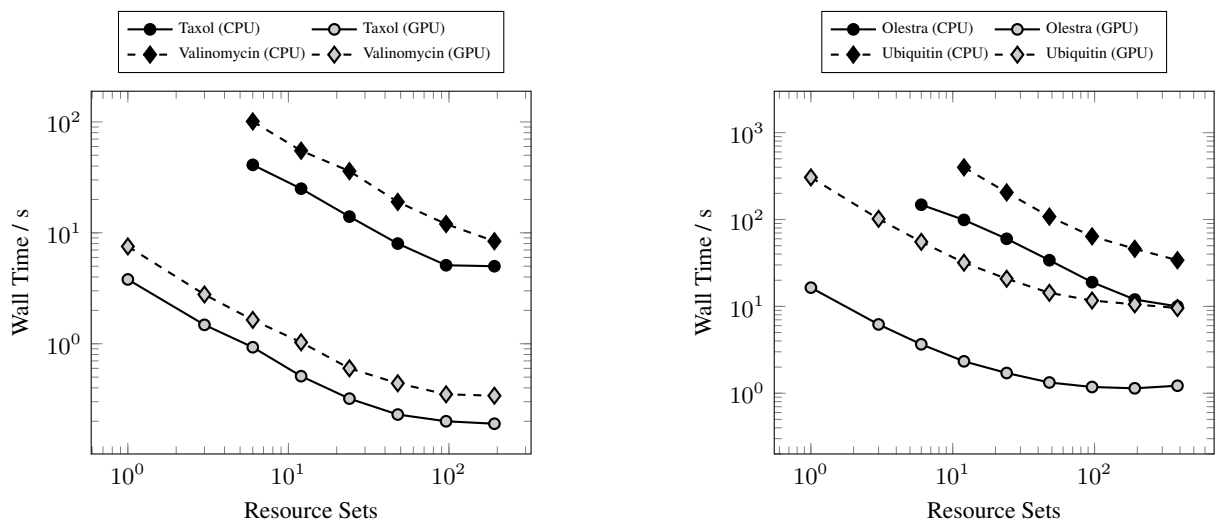
\begin{figure}[t]
  \centering
  \begin{minipage}{0.45\textwidth}
  \figname{small-molecule-strong-scaling}
  \begin{tikzpicture}

\begin{loglogaxis}[
  width=\textwidth,
  legend style={font=\tiny,at={(0.5,1.05)},anchor=south},
  legend cell align={left},
  legend columns=2,
  ticklabel style = {font=\small},
  label style = {font=\small},
  xlabel={Resource Sets},
  ylabel={Wall Time / s},
  cycle list={%
    {color=black, mark=*, mark options={solid}, thick},%
    {color=black, mark=*, mark options={solid,fill=black!20}, thick},%
    {dashed,color=black, mark=diamond*, mark options={solid,scale=1.5}, thick},%
    {dashed,color=black, mark=diamond*, mark options={solid,scale=1.5,fill=black!20}, thick},%
    {color=black, mark=diamond,  mark options={solid,scale=1.5}, thick},%
    {color=black, mark=diamond*, mark options={solid,scale=1.5}, thick},%
    {dashed,color=black, mark=o, mark options={solid}, thick},%
    {dashed,color=black, mark=*, mark options={solid}, thick}%
  }
]

\addlegendentry{Taxol (CPU)};
\addplot+ table [ x=RS, y=UFG-6-31G(d), col sep=comma ]
  {fig-src/data/taxol-nwchem.dat};

\addlegendentry{Taxol (GPU)};
\addplot+ table [ 
  x=RS, 
  y expr=\thisrow{Total} - \thisrow{ALLOCMax}, 
  col sep=comma 
] {fig-src/data/try3/taxol-6-31Gd-ufg.dat};

\addlegendentry{Valinomycin (CPU)};
\addplot+ table [ x=RS, y=UFG-6-31G(d), col sep=comma ]
  {fig-src/data/valinomycin-nwchem.dat};

\addlegendentry{Valinomycin (GPU)};
\addplot+ table [ 
  x=RS, 
  y expr=\thisrow{Total} - \thisrow{ALLOCMax}, 
  col sep=comma 
] {fig-src/data/try3/valinomycin-6-31Gd-ufg.dat};

\end{loglogaxis}

\end{tikzpicture}
  \end{minipage}~\hfill
  \begin{minipage}{0.45\textwidth}
  \figname{olestra-ubiquitin-strong-scaling}
  \begin{tikzpicture}

\begin{loglogaxis}[
  width=\textwidth,
  legend style={font=\tiny,at={(0.5,1.05)},anchor=south},
  legend cell align={left},
  legend columns=2,
  ticklabel style = {font=\small},
  label style = {font=\small},
  xlabel={Resource Sets},
  ylabel={Wall Time / s},
  ymin=0.2,ymax=3000,
  cycle list={%
    {color=black, mark=*, mark options={solid}, thick},%
    {color=black, mark=*, mark options={solid,fill=black!20}, thick},%
    {dashed,color=black, mark=diamond*, mark options={solid,scale=1.5}, thick},%
    {dashed,color=black, mark=diamond*, mark options={solid,scale=1.5,fill=black!20}, thick},%
    {color=black, mark=diamond,  mark options={solid,scale=1.5}, thick},%
    {color=black, mark=diamond*, mark options={solid,scale=1.5}, thick},%
    {dashed,color=black, mark=o, mark options={solid}, thick},%
    {dashed,color=black, mark=*, mark options={solid}, thick}%
  }
]

\addlegendentry{Olestra (CPU)};
\addplot+ table [ x=RS, y=UFG-6-31G(d), col sep=comma ]
  {fig-src/data/olestra-nwchem.dat};

\addlegendentry{Olestra (GPU)};
\addplot+ table [ 
  x=RS, 
  y expr=\thisrow{Total} - \thisrow{ALLOCMax}, 
  col sep=comma 
] {fig-src/data/try3/olestra-6-31Gd-ufg.dat};

\addlegendentry{Ubiquitin (CPU)};
\addplot+ table [ x=RS, y=UFG-6-31G(d), col sep=comma ]
  {fig-src/data/ubiquitin-nwchem.dat};

\addlegendentry{Ubiquitin (GPU)};
\addplot+ table [ 
  x=RS, 
  y expr=\thisrow{Total} - \thisrow{ALLOCMax}, 
  col sep=comma 
] {fig-src/data/try3/ubiquitin-6-31Gd-ufg.dat};

\end{loglogaxis}

\end{tikzpicture}
  \end{minipage}
  \caption{Strong scaling comparisons for the CPU (NWChem) and GPU (NWChemEx) 
  implementations of the XC integration. Timings for both NWChem and NWChemEx
  include all steps in the XC integration (batch generation, weight scaling,
  local integration and reduction).} 
  \label{fig:strong-scaling}
\end{figure}

\begin{figure}[t]
  \centering
  \begin{minipage}{0.45\textwidth}
  \figname{small-molecule-speedup}
  \begin{tikzpicture}

\begin{loglogaxis}[
  width=\textwidth,
  legend style={font=\tiny,at={(0.5,1.05)},anchor=south},
  legend cell align={left},
  legend columns=3,
  ticklabel style = {font=\small},
  label style = {font=\small},
  xlabel={Resource Sets},
  ylabel={Speedup},
  ymin=1, ymax=150,
  cycle list={%
    {color=black, mark=*, mark options={solid}, thick},%
    {color=black, solid, mark=diamond*,%
     mark options={solid,scale=1.5}, thick},%
    {color=black, mark=square*, mark options={solid}, thick},%
    {color=black, mark=*, mark options={solid,fill=black!30}, dashed, thick},%
    {color=black, solid, mark=diamond*,%
     mark options={solid,fill=black!30,scale=1.5}, dashed, thick},%
    {color=black, mark=square*, mark options={solid,fill=black!30},%
     dashed,thick}%
  }
]

\addlegendentry{Taxol (FG)};
\addplot+ table [ x=RS, y=6-31Gd-FG, col sep=comma ]
  {fig-src/data/try3/taxol-speedup.dat};

\addlegendentry{Taxol (UFG)};
\addplot+ table [ x=RS, y=6-31Gd-UFG, col sep=comma ]
  {fig-src/data/try3/taxol-speedup.dat};

\addlegendentry{Taxol (SFG)};
\addplot+ table [ x=RS, y=6-31Gd-SFG, col sep=comma ]
  {fig-src/data/try3/taxol-speedup.dat};

\addlegendentry{Valinomycin (FG)};
\addplot+ table [ x=RS, y=6-31Gd-FG, col sep=comma ]
  {fig-src/data/try3/valinomycin-speedup.dat};

\addlegendentry{Valinomycin (UFG)};
\addplot+ table [ x=RS, y=6-31Gd-UFG, col sep=comma ]
  {fig-src/data/try3/valinomycin-speedup.dat};

\addlegendentry{Valinomycin (SFG)};
\addplot+ table [ x=RS, y=6-31Gd-SFG, col sep=comma ]
  {fig-src/data/try3/valinomycin-speedup.dat};

\end{loglogaxis}

\end{tikzpicture}
  \end{minipage}~\hfill
  \begin{minipage}{0.45\textwidth}
  \figname{olestra-ubiquitin-speedup}
  \begin{tikzpicture}

\begin{loglogaxis}[
  width=\textwidth,
  legend style={font=\tiny,at={(0.5,1.05)},anchor=south},
  legend cell align={left},
  legend columns=3,
  ticklabel style = {font=\small},
  label style = {font=\small},
  xlabel={Resource Sets},
  ylabel={Speedup},
  ymin=1, ymax=150,
  cycle list={%
    {color=black, mark=*, mark options={solid}, thick},%
    {color=black, solid, mark=diamond*,%
     mark options={solid,scale=1.5}, thick},%
    {color=black, mark=square*, mark options={solid}, thick},%
    {color=black, mark=*, mark options={solid,fill=black!30}, dashed, thick},%
    {color=black, solid, mark=diamond*,%
     mark options={solid,fill=black!30,scale=1.5}, dashed, thick},%
    {color=black, mark=square*, mark options={solid,fill=black!30},%
     dashed,thick}%
  }
]

\addlegendentry{Olestra (FG)};
\addplot+ table [ x=RS, y=6-31Gd-FG, col sep=comma ]
  {fig-src/data/try3/olestra-speedup.dat};

\addlegendentry{Olestra (UFG)};
\addplot+ table [ x=RS, y=6-31Gd-UFG, col sep=comma ]
  {fig-src/data/try3/olestra-speedup.dat};

\addlegendentry{Olestra (SFG)};
\addplot+ table [ x=RS, y=6-31Gd-SFG, col sep=comma ]
  {fig-src/data/try3/olestra-speedup.dat};

\addlegendentry{Ubiquitin (FG)};
\addplot+ table [ x=RS, y=6-31Gd-FG, col sep=comma ]
  {fig-src/data/try3/ubiquitin-speedup.dat};

\addlegendentry{Ubiquitin (UFG)};
\addplot+ table [ x=RS, y=6-31Gd-UFG, col sep=comma ]
  {fig-src/data/try3/ubiquitin-speedup.dat};

\addlegendentry{Ubiquitin (SFG)};
\addplot+ table [ x=RS, y=6-31Gd-SFG, col sep=comma ]
  {fig-src/data/try3/ubiquitin-speedup.dat};

\end{loglogaxis}

\end{tikzpicture}
  \end{minipage}
  \caption{Achieved speedups of the GPU (NWChemEx) implementation over the CPU (NWChem)
  implementation of the XC integration for the 6-31G(d) basis set.}
  \label{fig:speedup}
\end{figure}
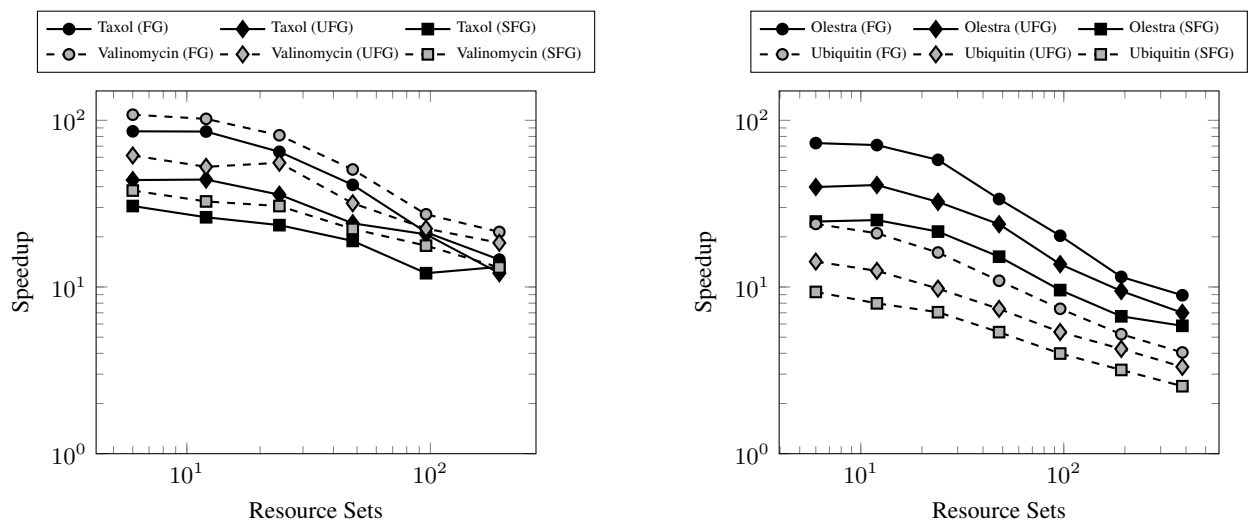

The primary goal of this work has been to provide a scalable implementation of
the XC integration. As such, we examine
the strong scaling the proposed method in comparison with the CPU
implementation in NWChem. Strong scaling results for the CPU and GPU XC
integrations using the 6-31G(d) basis and UFG quadrature are given in \cref{fig:strong-scaling}.
The wall times presented in \cref{fig:strong-scaling} only include those
operations that are required to perform the XC integration; wall times for the
allocation of device memory in the NWChemEx results have been removed.  
For the smallest problems (Taxol and Valinomycin), both NWChem and NWChemEx
exhibit near linear strong scaling out to 4 Summit nodes (168 MPI ranks in the case of
NWChem, and 24 GPUs in the case of NWChemEx). For largest problems (Olestra and
Ubiquitin), linear strong scaling is exhibited up to 8 Summit notes (48 GPUs)
in the case of NWChemEx and 16 nodes (336 MPI ranks) in the case of NWChem. The
relative speedups of NWChemEx over NWChem for the considered systems in the 6-31G(d) basis set 
is given in \cref{fig:speedup}. For
all but the largest problem (Ubiquitin), speedups over 10x are observed over
the CPU implementation at all resource set counts. 
For the smallest problems with the smallest grid size (FG), speedups of
$\sim$100x are observed when run on a small number of resource sets.
The degradation in speedup as a function in quadrature size is due to the
aforementioned differences in weight and density screening techniques between
NWChem and NWChemEx.  The magnitude of these speedups decrease as the amount of
resources increase.  This is especially the case for Ubiquitin, where a speedup
of $\sim$10x is observed at a single Summit node, but this speed up falls to
nearly 2.5x in the strong scaling limit. To better understand the stagnation of
strong scaling in this case, it is necessary to examine the scaling of the
individual components of the XC integration.

\begin{figure}[t]
  \centering
  \begin{minipage}{0.45\textwidth}
  \figname{valinomycin-component}
  \begin{tikzpicture}

\begin{loglogaxis}[
  width=\textwidth,
  legend style={font=\tiny},
  legend cell align={left},
  legend pos=north east,
  ticklabel style = {font=\small},
  label style = {font=\small},
  xlabel={Resource Sets},
  ylabel={Wall Time / s},
  xmin=1,xmax=192,
  cycle list={%
    {thick,black, mark=*,         mark options={solid}},%
    {thick,black, mark=square*,   mark options={solid,fill=black!30}},%
    {thick,black, mark=triangle*, mark options={solid,scale=1.5,fill=black!30}},%
    {thick,black, mark=diamond*,  mark options={solid,scale=1.5,fill=black!30}}%
  }
]

\addplot+[forget plot] 
plot  [error bars/.cd, y dir=plus, y explicit]
table [ x=RS, y=PROCESSAvg, y error expr=\thisrow{PROCESSMax}-\thisrow{PROCESSAvg}, col sep=comma ]
  {fig-src/data/try3/valinomycin-6-31Gd-ufg.dat};

\addplot+
plot  [error bars/.cd, y dir=minus, y explicit]
table [ x=RS, y=PROCESSAvg, y error expr=\thisrow{PROCESSAvg}-\thisrow{PROCESSMin}, col sep=comma ]
  {fig-src/data/try3/valinomycin-6-31Gd-ufg.dat};

\addplot+[forget plot] 
plot  [error bars/.cd, y dir=plus, y explicit]
table [ x=RS, y=REDUCEAvg, y error expr=\thisrow{REDUCEMax}-\thisrow{REDUCEAvg}, col sep=comma, restrict expr to domain={\thisrow{RS}}{3:384} ]
  {fig-src/data/try3/valinomycin-6-31Gd-ufg.dat};

\addplot+ 
plot  [error bars/.cd, y dir=minus, y explicit]
table [ x=RS, y=REDUCEAvg, y error expr=\thisrow{REDUCEAvg}-\thisrow{REDUCEMin}, col sep=comma, restrict expr to domain={\thisrow{RS}}{3:384} ]
  {fig-src/data/try3/valinomycin-6-31Gd-ufg.dat};

\addplot+[forget plot] 
plot  [error bars/.cd, y dir=plus, y explicit]
table [ x=RS, y=BATCHAvg, y error expr=\thisrow{BATCHMax}-\thisrow{BATCHAvg}, col sep=comma ]
  {fig-src/data/try3/valinomycin-6-31Gd-ufg.dat};

\addplot+
plot  [error bars/.cd, y dir=minus, y explicit]
table [ x=RS, y=BATCHAvg, y error expr=\thisrow{BATCHAvg}-\thisrow{BATCHMin}, col sep=comma ]
  {fig-src/data/try3/valinomycin-6-31Gd-ufg.dat};

\addplot+ table [ 
  x=RS, 
  y expr=\thisrow{Total} - \thisrow{ALLOCMax}, 
  col sep=comma 
] {fig-src/data/try3/valinomycin-6-31Gd-ufg.dat};

\draw[dashed,thick] (1,5) -- (192, 5/192);

\end{loglogaxis}
\node at (0.5,0.3) {(a)};

\end{tikzpicture}
  \end{minipage}~\hfill
  \begin{minipage}{0.45\textwidth}
  \figname{ubiquitin-component}
  \begin{tikzpicture}

\begin{loglogaxis}[
  width=\textwidth,
  legend style={font=\tiny,draw=none},
  legend cell align={left},
  legend pos=north east,
  ticklabel style = {font=\small},
  label style = {font=\small},
  xlabel={Resource Sets},
  ylabel={Wall Time / s},
  xmin=1,xmax=384,
  cycle list={%
    {thick,black, mark=*,         mark options={solid}},%
    {thick,black, mark=square*,   mark options={solid,fill=black!30}},%
    {thick,black, mark=triangle*, mark options={solid,scale=1.5,fill=black!30}},%
    {thick,black, mark=diamond*,  mark options={solid,scale=1.5,fill=black!30}}%
  }
]

\addlegendentry{Local Integration};
\addplot+[forget plot] 
plot  [error bars/.cd, y dir=plus, y explicit]
table [ x=RS, y=PROCESSAvg, y error expr=\thisrow{PROCESSMax}-\thisrow{PROCESSAvg}, col sep=comma ]
  {fig-src/data/try3/ubiquitin-6-31Gd-ufg.dat};
\addplot+
plot  [error bars/.cd, y dir=minus, y explicit]
table [ x=RS, y=PROCESSAvg, y error expr=\thisrow{PROCESSAvg}-\thisrow{PROCESSMin}, col sep=comma ]
  {fig-src/data/try3/ubiquitin-6-31Gd-ufg.dat};

\addlegendentry{Reduce};
\addplot+[forget plot] 
plot  [error bars/.cd, y dir=plus, y explicit]
table [ x=RS, y=REDUCEAvg, y error expr=\thisrow{REDUCEMax}-\thisrow{REDUCEAvg}, col sep=comma, restrict expr to domain={\thisrow{RS}}{3:400} ]
  {fig-src/data/try3/ubiquitin-6-31Gd-ufg.dat};
\addplot+
plot  [error bars/.cd, y dir=minus, y explicit]
table [ x=RS, y=REDUCEAvg, y error expr=\thisrow{REDUCEAvg}-\thisrow{REDUCEMin}, col sep=comma, restrict expr to domain={\thisrow{RS}}{3:400} ]
  {fig-src/data/try3/ubiquitin-6-31Gd-ufg.dat};

\addlegendentry{Load Balance};
\addplot+[forget plot] 
plot  [error bars/.cd, y dir=plus, y explicit]
table [ x=RS, y=BATCHAvg, y error expr=\thisrow{BATCHMax}-\thisrow{BATCHAvg}, col sep=comma ]
  {fig-src/data/try3/ubiquitin-6-31Gd-ufg.dat};
\addplot+
plot  [error bars/.cd, y dir=minus, y explicit]
table [ x=RS, y=BATCHAvg, y error expr=\thisrow{BATCHAvg}-\thisrow{BATCHMin}, col sep=comma ]
  {fig-src/data/try3/ubiquitin-6-31Gd-ufg.dat};

\addlegendentry{Total Time};
\addplot+ table [ 
  x=RS, 
  y expr=\thisrow{Total} - \thisrow{ALLOCMax}, 
  col sep=comma 
] {fig-src/data/try3/ubiquitin-6-31Gd-ufg.dat};

\addlegendentry{Linear};
\addplot+[black,dashed,thick, no markers] table [ x=x, y=y ] {
  x   y
  1   200
  384 0.52083333333
};

\end{loglogaxis}
\node at (0.5,0.3) {(b)};

\end{tikzpicture}
  \end{minipage}~\hfill
  \caption{Strong scaling of individual components of the XC Integration 
  for Valinomycin (a) and Ubiquitin (b) in
  comparison to total execution time.  Error bars represent min/max times and
  solid markers represent average wall time over all resource sets.}
  \label{fig:component-ufg}
\end{figure}
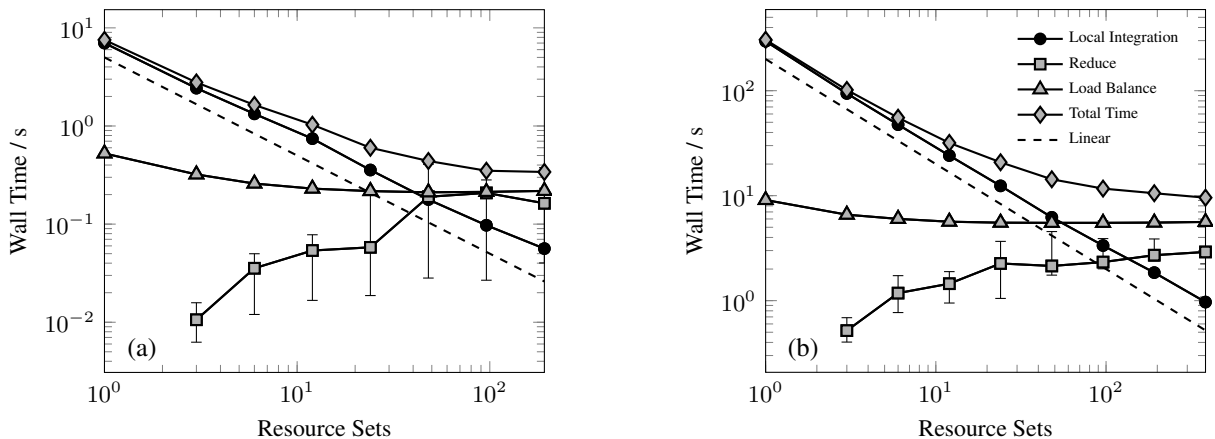

\Cref{fig:component-ufg} shows the timings for various components of the GPU
XC integration for considered systems. Rather than examine the scaling for each
of the considered systems, we choose to profile the largest of the small sized
problems (Valinomycin), and the largest problem (Ubiquitin) as
representative test cases. 
%
%
%
As can be seen in \cref{fig:component-ufg}, the local integration scales
linearly for all processor counts considered.  As the local
integration scales linearly,  stagnation is not due to a lack of sufficient
work to occupy the GPU, but rather due to the increasing cost of the MPI
reduction and the constant cost of replicating \cref{alg:load_balance} on
all resource sets. 
This scaling behavior could be further improved by
porting \cref{alg:load_balance} to the GPU, however, in the case of large
processor counts, the reduction becomes competitive with
\cref{alg:load_balance}, thus it would be unlikely to demonstrate any
qualitatively different scaling behavior in this regime.

\section{Conclusion}
\label{sec:conclusions}

In this work, we have proposed and implemented a three-level, GPU-based  parallelism scheme for
the distributed numerical integration of the XC potential and energy required
for the evaluation of the Fock matrix in the Gaussian basis discretization of
KS-DFT. In addition to the development of a simple load balancing scheme, the
method proposed in this work for the evaluation of local integration quantities
emphasizes the use of batched kernel invocations to achieve high throughput
in the evaluation of localized quadrature batches on the GPU. This
approach was motivated by the recent advent of GPU accelerated batched BLAS
kernels which have seen wide adoption in many GPU applications. We have
demonstrated the proposed load balancing scheme produces linear strong scaling
in the local integration of XC quantities for the problems considered. 
Further we have validated the efficacy of the use of batched kernels, including
the use of batched GEMM and SYR2K, by demonstrating the ability of these
kernels to achieve excellent efficiency on the NVIDIA Tesla V100 for a
wide range of molecular systems, basis sets and quadrature sizes. 

The largest deficiency in the current work is the restricted implementation of
the GPU related techniques to NVIDIA GPUs and the CUDA SDK. As of this work,
emerging architectures are increasingly relying upon other GPU vendors (AMD,
Intel, etc) which would render direct application of the current implementation
impossible.  However, the principles of batched kernel evaluation may be
extended to many if not all GPU devices. Thus, as has been explored in the
context of related implementations of seminumerical exchange calculations
\cite{laqua20_highly}, future work will focus on the \emph{portable}
implementation of the scalable GPU method presented in this work.

We have implemented the proposed method in the open-source NWChemEx software
package and have demonstrated speedups between 10x-100x over the analogous CPU
implementation in NWChem. However, in the strong scaling limit, the proposed
replicated load balance scheme and distributed reduction of XC integrands
become computationally dominant which causes early stagnation relative to
the linearly scaling local integration on the GPU. As has been demonstrated
in related work \cite{madushanka20_parallel}, porting the batch generation and screening
procedure to the GPU would help mitigate the strong scaling stagnation, though
the asymptotic bottleneck of the distributed reduction would still remain.
With the one-to-one CPU-to-GPU affinity discussed in this work, the
computational cost of the MPI reduction could be reduced through the use of
remote memory access (RMA) to exploit shared memory spaces and void explicit
data communication.  As the local integration scales linearly out to very large
processor and GPU counts, further improvements in these non-GPU aspects of the XC
integration would drastically improve the strong scaling of the proposed
methods. Such improvements will be explored in future work.

\section{Acknowledgments}

The authors would also like to thank Eduardo Aprà and Ajay Panyala of the
Pacific Northwest National Laboratory (PNNL) for insightful discussions
regarding DFT calculations and code modifications related to the NWChem
software package to enable meaningful comparisons with the methods presented in
this work.
This research was supported by the Exascale Computing Project (17-SC-20-SC), a
collaborative effort of the U.S. Department of Energy Office of Science and the
National Nuclear Security Administration. This research used resources of the
Oak Ridge Leadership Computing Facility, which is a DOE Office of Science User
Facility supported under Contract DE-AC05-00OR22725.


\ifarxivtypeset
  \bibliographystyle{acm} 
\else
  \bibliographystyle{frontiersinSCNS_ENG_HUMS} 
\fi
\bibliography{refs, MAGMA_Publications.bib}

\end{document}